\documentclass{ws-p8-50x6-00}
\newcommand{\unit}[1]{\,\mathrm{#1}}

\begin{document}
\newcommand{\M}{\mbox{m}}
\newcommand{\n}{\mbox{$n_f$}}
\newcommand{\EP}{\mbox{e$^+$}}
\newcommand{\EM}{\mbox{e$^-$}}
\newcommand{\EPEM}{\mbox{e$^+$e$^-$}}
\newcommand{\EMEM}{\mbox{e$^-$e$^-$}}
\newcommand{\GG}{\mbox{$\gamma\gamma$}}
\newcommand{\GE}{\mbox{$\gamma$e}}
\newcommand{\EE}{\mbox{e$^-$e$^-$}}
\newcommand{\GP}{\mbox{$\gamma$e$^+$}}
\newcommand{\TEV}{\mbox{TeV}}
\newcommand{\GEV}{\mbox{GeV}}
\newcommand{\LGG}{\mbox{$L_{\gamma\gamma}$}}
\newcommand{\LGE}{\mbox{$L_{\gamma e}$}}
\newcommand{\LEE}{\mbox{$L_{ee}$}}
\newcommand{\WGG}{\mbox{$W_{\gamma\gamma}$}}
\newcommand{\EV}{\mbox{eV}}
\newcommand{\CM}{\mbox{cm}}
\newcommand{\MM}{\mbox{mm}}
\newcommand{\NM}{\mbox{nm}}
\newcommand{\MKM}{\mbox{$\mu$m}}
\newcommand{\SEC}{\mbox{s}}
\newcommand{\CMS}{\mbox{cm$^{-2}$s$^{-1}$}}
\newcommand{\MRAD}{\mbox{mrad}}
\newcommand{\IND}{\hspace*{\parindent}}
\newcommand{\E}{\mbox{$\epilon$}}
\newcommand{\EN}{\mbox{$\epsilon_n$}}
\newcommand{\EI}{\mbox{$\epsilon_i$}}
\newcommand{\ENI}{\mbox{$\epsilon_{ni}$}}
\newcommand{\ENX}{\mbox{$\epsilon_{nx}$}}
\newcommand{\ENY}{\mbox{$\epsilon_{ny}$}}
\newcommand{\EX}{\mbox{$\epsilon_x$}}
\newcommand{\EY}{\mbox{$\epsilon_y$}}  
\newcommand{\BI}{\mbox{$\beta_i$}}
\newcommand{\BX}{\mbox{$\beta_x$}}
\newcommand{\BY}{\mbox{$\beta_y$}}
\newcommand{\SX}{\mbox{$\sigma_x$}}
\newcommand{\SY}{\mbox{$\sigma_y$}}
\newcommand{\SZ}{\mbox{$\sigma_z$}}
\newcommand{\SI}{\mbox{$\sigma_i$}}
\newcommand{\SIP}{\mbox{$\sigma_i^{\prime}$}}
\newcommand{\NC}{\mbox{${\cal NC}$}}
\newcommand{\CC}{\mbox{${\cal CC}$}}
\def\b{\beta}
\def\g{\gamma}
\def\SM{$\mathcal{SM}$}
\def\MSSM{$\mathcal{MSSM}$}
\def\2HDM{$2\mathcal{HDM}$}
\def\h{\rm h}
\def\ccbar{\overline{\mbox c}\mbox{c}}
\def\bbbar{\overline{\mbox b}\mbox{b}}
\def\qqbar{\overline{\mbox q}\mbox{q}}
\def\ccbarg{\overline{\mbox c}\mbox{cg}}
\def\bbbarg{\overline{\mbox b}\mbox{bg}}
\def\BR{\rm BR}
\newcommand{\bear}{\begin{equation}\begin{array}}

\newcommand{\sw}{\mbox{$\sin\Theta_W\,$}}

\newcommand{\cw}{\mbox{$\cos\Theta_W\,$}}
\newcommand{\epe}{\mbox{$e^+e^-\,$}}
\newcommand{\ggam}{\mbox{$\gamma\gamma\,$}}
\newcommand{\egam}{\mbox{$e\gamma\,$}}
\newcommand{\gewnu}{\mbox{$e\gamma\to W\nu\,$}}
\newcommand{\eeww}{\mbox{$e^+e^-\to W^+W^-\,$}}
\newcommand{\ggww}{\mbox{$\gamma\gamma\to W^+W^-\,$}}
\newcommand{\ggzz}{\mbox{$\gamma\gamma\to ZZ\,$}}
\newcommand{\egeh}{\mbox{$e\gamma\to eH\,$}}
\newcommand{\geeww}{\mbox{$e\gamma\to e W^+W^-\,$}}
\newcommand{\beq}{\begin{equation}}
\newcommand{\eeq}{\end{equation}}
\newcommand{\beqn}{\begin{eqnarray}}
\newcommand{\eeqn}{\end{eqnarray}}
\newcommand{\lum}[1]{{\rm luminosity} $ #1 $ cm$^{-2}$ s$^{-1}\,$}
\newcommand{\intlum}[1]{{\rm annual luminosity} $ #1$  fb$^{-1}\,$}
\newcommand{\mw}{\mbox{$M_W\,$}}
\newcommand{\mww}{\mbox{$M_W^2\,$}}
\newcommand{\mh}{\mbox{$M_H\,$}}
\newcommand{\mhh}{\mbox{$M_H^2\,$}}
\newcommand{\mz}{\mbox{$M_Z\,$}}
\newcommand{\mzz}{\mbox{$M_Z^2\,$}}
\newcommand{\sigmaw}{\mbox{$\sigma_W\,$}}
\newcommand{\sww}{\mbox{$\sin^2\Theta_W\,$}}
\newcommand{\cww}{\mbox{$\cos^2\Theta_W\,$}}
\newcommand{\ptr}{\mbox{$p_{\bot}\,$}}
\newcommand{\ptrs}{\mbox{$p_{\bot}^2\,$}}
\newcommand{\lgam}{\mbox{$\lambda_{\gamma}$}}
\newcommand{\lga}[1]{\mbox{$\lambda_{#1}$}}
\newcommand{\lggam}[2]{\mbox{$\lambda_{#1}\lambda_{#2}$}}
\newcommand{\lgg}{\lambda_1\lambda_2}
\newcommand{\lel}{\mbox{$\lambda_e$}}
\newcommand{\ggh}{\mbox{$\gamma\gamma\to hadrons$}}
\newcommand{\pair}[1]{\mbox{$#1 \bar{#1}$}}
\newcommand{\ZZ}{\mbox{ZZ}}
\newcommand{\WW}{\mbox{WW}}
\newcommand{\Z}{\mbox{Z}}
\newcommand{\W}{\mbox{W}}

\title{Summary of Photon 2001}

\author{Armin B\"ohrer}

\address{Fachbereich Physik, Universit\"at Siegen, 57068 Siegen, Germany\\
E-mail: armin.boehrer@cern.ch}

\author{Maria Krawczyk}

\address{IFT UW, Hoza 69, 00-681 Warszawa, Poland\\
E-mail: Maria.Krawczyk@fuw.edu.pl}


\maketitle

\abstracts{This article summarizes the experimental and theoretical 
results presented and discussed  at the International Conference 
on the Structure and Interactions of the Photon, 
PHOTON 2001, in Ascona.}

\section{Introduction}
 The photon, the first and best known gauge boson, is a fundamental
 particle with  pointlike  couplings to  charged fundamental particles.
  In  high energy hadronic processes, however, the photon may behave 
  like a hadron. 
The study of the structure and the interaction of the 
 photon in soft, semi-hard and hard hadronic processes 
 turns out to be a rich field of investigation and test of the 
strong interaction. Detailed studies are performed especially at 
 ${\mathrm {e^+ e^-}}$ colliders 
 and now also with ${\mathrm {ep}}$ interactions. 
 Different in their initial state, processes at LEP 
and HERA add to give a common understanding of the photon and its 
interaction. This is rounded off by specific, detailed and complementary 
results at low energy  colliders also discussed at this conference.

In this summary we present the highlights of the conference PHOTON 2001 as 
seen by the two authors.
For further details we refer the reader to the individual presentations 
or the group summaries of the conference and references 
therein.
\section{Theoretical Framework}
The {\it soft}  high-energy  processes involving 
photons, a bulk of events, can be described in the  
VDM-type models using   a framework of Regge Theory,
 while in the {\it hard} 
  processes the perturbative QCD works; the {\it semi-hard} processes 
 corresponds to 
 the case where both Regge approach and pQCD  are applicable.
In both,  hard 
and semi-hard  processes the notion of the partonic structure 
of the photon is useful and it is clear that in such processes the photon 
may participate directly or via its partonic agents (the resolved photon 
processes).  

The standard DGLAP approach  to describe  hard processes involving photons, 
like  DIS on the photon  and large $p_T$ jet (particle) production
in $\gamma \gamma$ or $\gamma {\mathrm p}$ collision, 
is based on the collinear approximation where 
terms with large $\log Q^2$ or $\log p_T^2$ are summed up.  
The leading and next-to-leading (LO and NLO) 
calculations exists for many quantities, 
also all existing parton parametrizations for the photon 
were obtained in such an approach. 

For semi-hard processes, e.g., DIS at  small $x$, 
there appear other large scale and corresponding large logarithms
have to be summed to all orders.
Resummation of large $\log 1/x$ terms  can be obtained 
using  BFKL or CCFM evolution equations. In both of the latter approaches
the $p_t$ of the initial parton cascade is not ordered; the basic ingredients are the unintegrated parton densities to be convoluted with appropriate matrix elements with the initial parton being off-shell.
 These   approaches were presented during this conference, being superior 
in some cases to the standard DGLAP approach in the description of the data.

In many respects the approaches used to describe photons are similar
to ones used for a hadron, e.g., proton. One should be aware, however, that the possible direct coupling to a pointlike parton influences significantly 
 the whole picture. The Parton Model prediction can be derived for 
$F_2^{\gamma}$ from the basic process $\gamma^* \gamma \rightarrow 
{\mathrm {q \bar q}}$ giving a logarithmic dependence on $Q^2$. 
The corresponding DGLAP evolution equations, which take into account
additional  QCD radiation processes  are inhomogeneous, 
and in principle (unfortunately not in practice due to appearance of
 singularities) can be solved without any input ({\it asymptotic solution}).
Basic ideas and results can be found in recent reviews 
~\cite{proceeding1,proceeding3,generalstrfct}.


\section{Structure Functions}
The (hadronic) structure function $F_2^{\gamma}$ of the (real) photon 
is measured in the hadron production processes in 
unpolarized ${{\mathrm e}^+{\mathrm e}^-}$ collisions 
from the single-tagged events. The negative momentum transfer squared  
of the photon emitted by the tagged 
lepton is equal to $q^2 = -2 E_{\mathrm{tag}} E_{\mathrm{beam}} 
(1-\cos \theta)$. The second lepton stays undetected so the corresponding 
momentum transfer squared  $p^2=-P^2$ is small.
The process then can be viewed as a deep inelastic electron-photon scattering
 (DIS$_{\mathrm e\gamma}$), where a quasi-real photon (with ``mass squared'' $p^2$)
 is probed by a virtual photon with high virtuality  $Q^2=-q^2$. The 
differential cross section can be expressed as 
${\mathrm d^2}\sigma_{{\mathrm e}\gamma \rightarrow {\mathrm {eX}}} / 
{\mathrm d}x{\mathrm d}Q^2 = 
[( 1+ (1-y)^2 ) F_2^{\gamma}(x,Q^2) - y^2 F_{L}^{\gamma}(x,Q^2)] 
2\pi \alpha^2 / xQ^2$ 
with 
$x \approx Q^2 / (Q^2+W^2)$ and 
the inelasticity 
$y \approx 1 - (E_{\mathrm {tag}}/E_{\mathrm {beam}}) 
\cos ^2 \theta_{\mathrm {tag}}$. 

In most cases $y$ is small and the term with $F_L^{\gamma}$ can  be neglected.
Some experiments correct their results for it, with the increased 
precision of the measurement the correction should become mandatory, as it
 was  noted at the conference.
The  ${\mathrm e}\gamma$ cross section given above is used 
to extract the structure function $F_2^{\gamma}(x, Q^2)$
primarily related to the quark densities in the photon $q^{\gamma}(x,Q^2)$, 
see Ref.~\cite{nisius} and Ref.~\cite{generalstrfct}.

The measurements of $F^{\gamma}_2$ at LEP reached   $x$ as low as $10^{-3}$ 
and cover a range of $\langle Q^2 \rangle$ from 
about $2 \unit{GeV^2}$ to more than 
$750 \unit{GeV^2}$ (even above 1000 $\unit{GeV^2}$ for $Q^2$-dependence). 
All existing   data  are shown in  Fig.~\ref{nisius-figsum} as a function of 
$x$, the new data obtained at the highest $Q^2$  
by OPAL and DELPHI groups~\cite{taylor,tyapkin} are shown in the last diagram. 
 The logarithmic rise with $Q^2$ seen in data
  is  predicted by the theory. It appears 
 already in the Parton Model - QCD modifies this rise softly (logarithmically).
In Fig.~\ref{tyapkin-csilling} the evolution with $Q^2$ is shown for various
 $x$ regions (DELPHI) in comparison with other data, MC predictions 
and with few parametrizations. 
  The data (Fig.~\ref{nisius-figsum}) 
 are nicely described by the GRV-HO and SaS-1D parton 
 parameterizations for a real photon; with the present precision the $Q^2$ 
 dependence data (Fig.~\ref{tyapkin-csilling})
  start to challenge existing parton parametrizations and MC models. 
  Note, that all parametrizations for photon 
were  obtained from  data  collected before 1997.
The full  present set of over two hundred data points  
is being used in a new LO fit to    $F_2^{\gamma}$ 
based on DGLAP evolution equation~\cite{jankowski}.
The updated  parton parametrization for a real photon  can be used to describe
hard processes induced by the real photons at LEP and HERA, where some 
discrepancies are being  observed (see Sec.~3 and 4). This new parametrization
 will be important  also for studies of potential of  
future linear colliders, especially the photon colliders (see Sec.~7).\ 
\begin{figure}[t]
\epsfxsize=0.88\textwidth
\epsfysize=0.9\textwidth 
\epsfbox{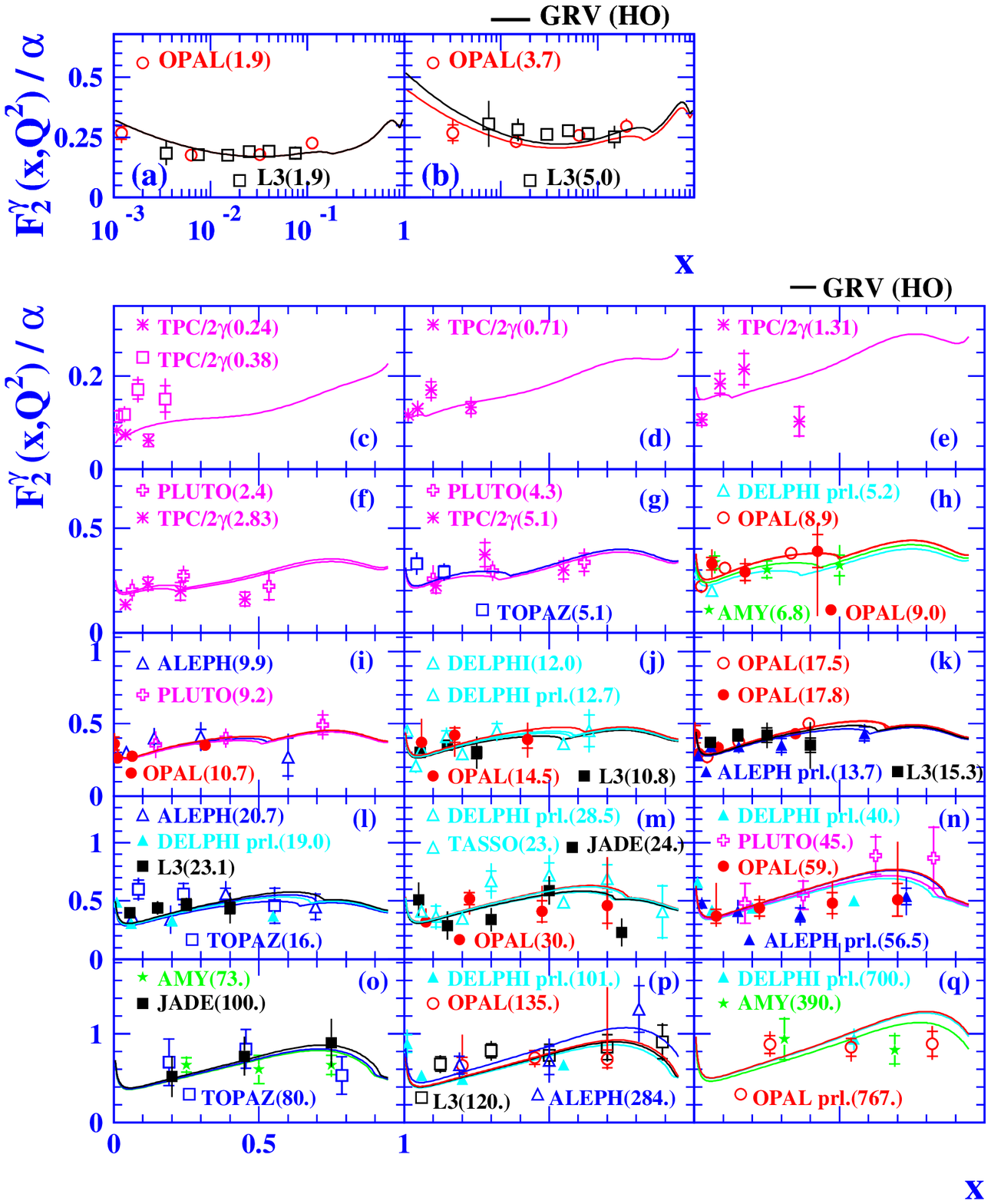}
\caption{Summary of measurements of the hadronic structure function 
$F_2^{\gamma}$, from~\protect\cite{nisius}.\label{nisius-figsum}}
\end{figure}

\begin{figure}[t]
\begin{minipage}{.50\textwidth}
\epsfxsize=0.97\textwidth
\epsfysize=0.97\textwidth 
\epsfbox{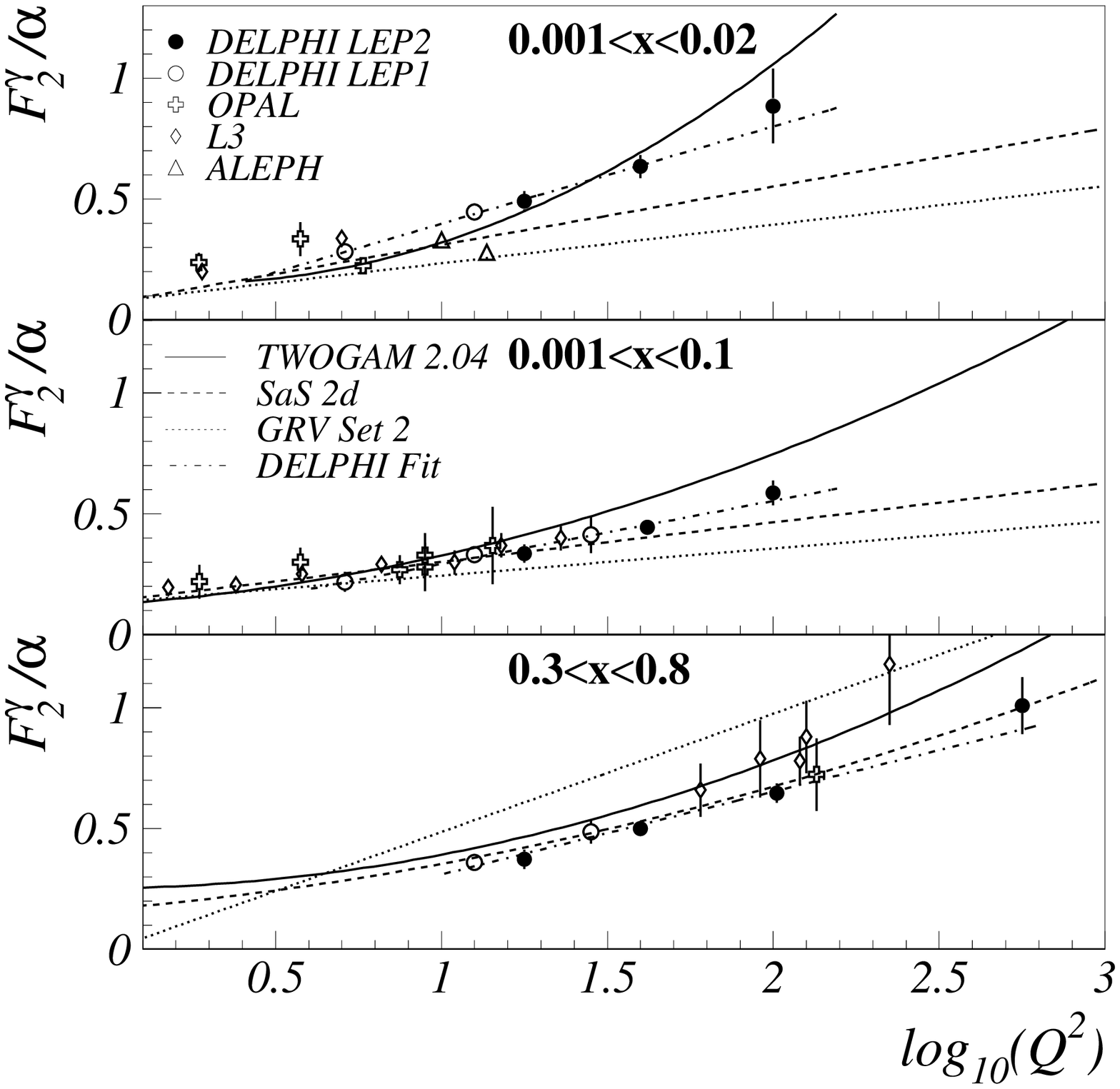}
\caption{Evolution of $F_2^{\gamma}$ with $Q^2$ in  various 
$x$- regions~\protect\cite{tyapkin}.}
\end{minipage}
~\
\begin{minipage}{.50\textwidth}
\epsfxsize=0.9\textwidth 
\epsfysize=0.9\textwidth 
\epsfbox{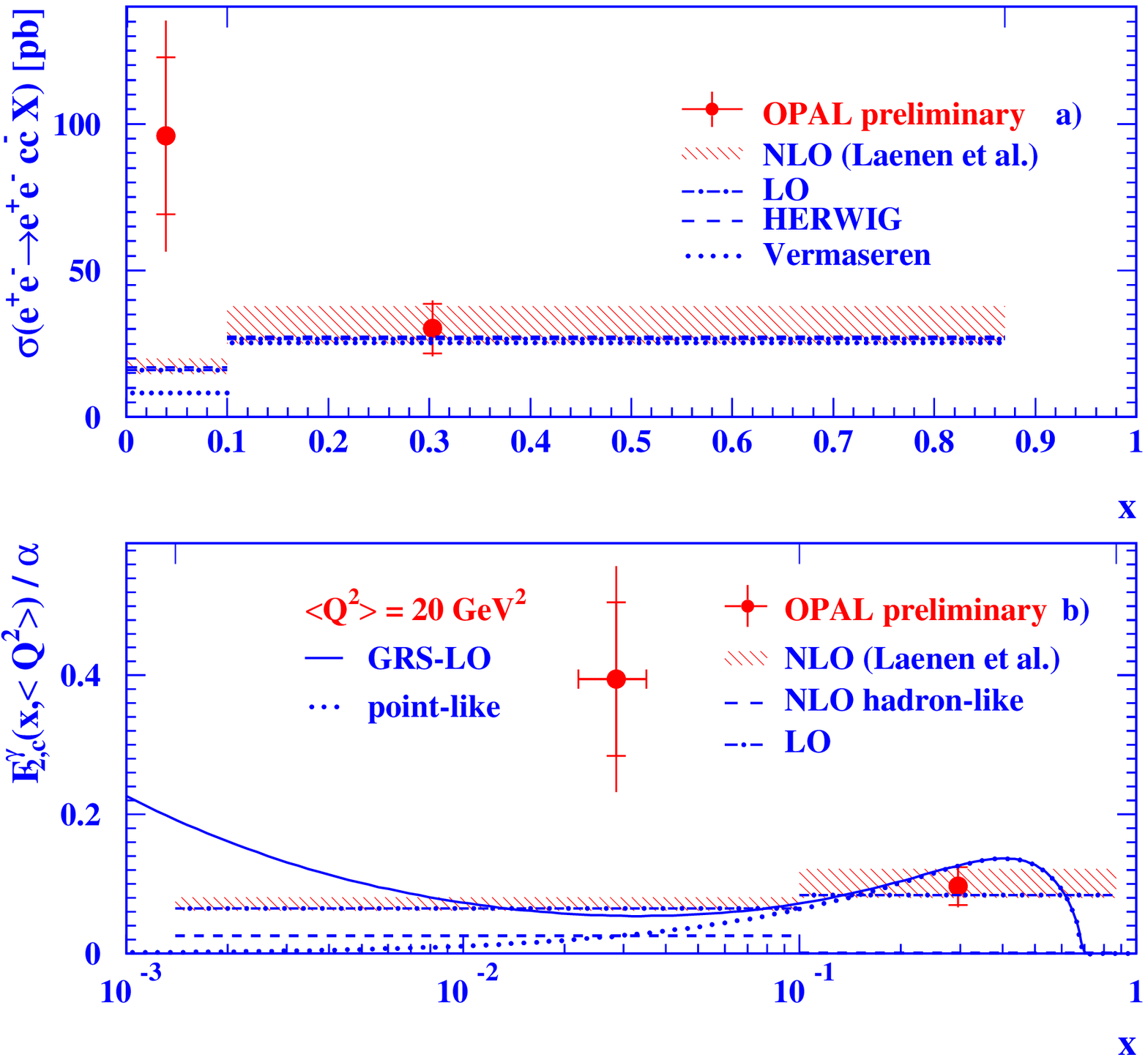} 
\caption{ Results for $F_{2,c}^{\gamma}$ 
\protect\cite{csilling}.
\label{tyapkin-csilling}}
\end{minipage}
\end{figure}

An extraction of $F_2^{\gamma}$ by fitting 
its components (QPM, VDM and RPC-resolved photon contribution) separately, 
rather 
than a regularized unfolding, may lead to a better measurement; the 
estimated uncertainties increase but are more reliable, as 
stated by DELPHI~\cite{tyapkin}. 
It has been pointed out at the conference, however, that improvements are 
still to be made: where precision is high by including the $P^2$ dependence 
or statistics is low by combining all four LEP experiments.
Note that if $P^2$ differs from zero, we deal with a virtual photon, which has 
  the three states of  polarization: two transverse(like a real photon) 
and one longitudinal (called also scalar) ones,
see Sec. 3.1 where the importance of the 
resolved $\gamma_L^*$ contribution are discussed. 

The charm contribution to the photon structure function $F_2^{\gamma}$, 
denoted $F_{2,{\mathrm c}}^{\gamma}$, can be obtained from the DIS$_{\mathrm e\gamma}$ 
data from the sample  with  charmed hadrons. The updated result of OPAL, 
which identifies charm by exclusive reconstruction of 
a D$^*$ meson (see also below), strengthens  their previous 
finding that  for $x < 0.1$ 
the predicted hadron-like contribution is lower than the 
measurement~\cite{csilling}, see Fig.~\ref{tyapkin-csilling}. 
A confirmation of the difference by the other LEP experiments is 
desirable. More data are needed to constrain the gluonic content of the 
photon in such measurements.

A first measurement has also been performed by OPAL of the process 
${\mathrm e}\gamma \rightarrow {\mathrm {eZ/\gamma^*}}$ where 
the centre-of-mass energy of the ${eZ/\gamma^*}$-system 
$\sqrt{\hat{s}}$ is equal or larger than the Z mass~\cite{fleck} 
and found to be in agreement with MC predictions.

\section{Hard Inclusive Processes}

Hard production of jets (single jet and dijets)
or particles  at LEP and HERA are  used as a complementary method to 
probe the ``structure'' of the photon~\cite{wing}. The hard scale $\tilde Q$
is then provided by a large $p_T$ of produced  jets or particles. 
The  part of momentum carried by 
a partonic constituent of the photon, $x_{\gamma}$, is at LO equal to $x$. 
Direct, single-resolved (for LEP also double-resolved) photon processes 
contribute, and can be separated at LO
by  using the fact that
in the resolved one the remnant jet carries away a part of the invariant
mass available in the $\gamma\gamma$  or $\gamma p$ collision.
For dijets  at HERA 
the variable $x_{\gamma}$,  can be reconstructed from
 the pseudorapidity $\eta$ and transverse 
momentum $E_T$ of the jets, 
as $x_{\gamma} = (E_T^{\mathrm {jet1}} {\mathrm e}^{- \eta^{\mathrm {jet1}}} 
+  E_T^{\mathrm {jet2}} {\mathrm e}^{- \eta^{\mathrm {jet2}}}) 
/ (2yE_{\mathrm e})$. Similarly $x_{\gamma}^{\pm}$ 
relevant for LEP measurements (photon-photon collisions) can be obtained.

The implementation of NLO in the event simulation for jet production 
is in progress as was reported at this conference, where a systematic 
method for combining NLO QCD calculation with the parton was 
presented~\cite{potter}.
{Special emphasis was recently put on the 
study of the (soft) underlying events (sue), especially important 
for the double-resolved contribution.  Inclusion of multiple 
parton interaction (MIA), parametrized as in  pure hh collision,  
 usually  improves  the agreement with data.}

The structure of jets, forming subjets, has been 
investigated at HERA~\cite{vasquez}, where samples of different 
gluon purity have been obtained in events with charm quarks. The 
subjet structure is predicted by QCD as a function of the jet resolution 
parameter. The results are consistent with perturbative QCD and 
consistent with findings in hadronic Z decays at LEP. 
Further tests of QCD such as the measurement of $\alpha_{\mathrm S}$ 
give values compatible 
in precision with other measurements or better~\cite{schoerner}.

Both at LEP (see Ref.~\cite{generalincl} for reviews) 
and HERA inclusive hadron  production has been studied intensively. 
Predictions from MLLA, e.g., average charged multiplicity, 
Gaussian shape of the 
distribution in $\xi = -\ln (p_{\mathrm {hadron}}/p_{\mathrm {beam}})$ 
and the dependence of the maximum $\xi^*$ on $Q^2$, were shown at 
this conference for HERA 
and found in very good agreement 
with the data both in deep inelastic and diffractive scattering 
processes, respectively, proving also the 
universality of the fragmentation functions. Such studies were performed for 
charged particles inclusively, for strange particles and particles 
versus anti-particles~\cite{traynorboogert} (see also Ref.~\cite{padhi} 
for similar studies in charmed jets at HERA).

\subsection{Jets}

$\bullet${ Dijets at LEP}\\

Present measurements  of dijet events at LEP, 
from a $\sim$ 600 pb$^{-1}$ data sample  collected,
 concentrate on the separation 
of the direct and (single)-resolved contribution to investigate the gluon 
content of the real photon~\cite{wengler,masik}, 
which is not tightly constrained by the $F_2^{\gamma}$ data.

Comparisons with the NLO calculations and the Monte Carlo models 
PYTHIA and HERWIG are made by OPAL  for $E_T$, and $\eta$  
as well as  $x_{\gamma}^{\pm}$ distributions 
 (see Fig.~\ref{wengler-fig2}). The MC predictions for both  
 parton parametrizations, GRS or SaS-1D, are  too low by 
20\% for small $x_{\gamma}$ and small $E_T$ indicating a too 
low gluon content in the photon. To study the effects of 
the underlying event 
the data sample is divided into samples where 
$x_{\gamma}^+$ or $x_{\gamma}^- < 0.75$ and $>0.75$, giving data sets with different 
single- and double-resolved fraction.
\begin{figure}[h]
\epsfxsize=0.50\textwidth
\epsfysize=0.58\textwidth 
\epsfbox{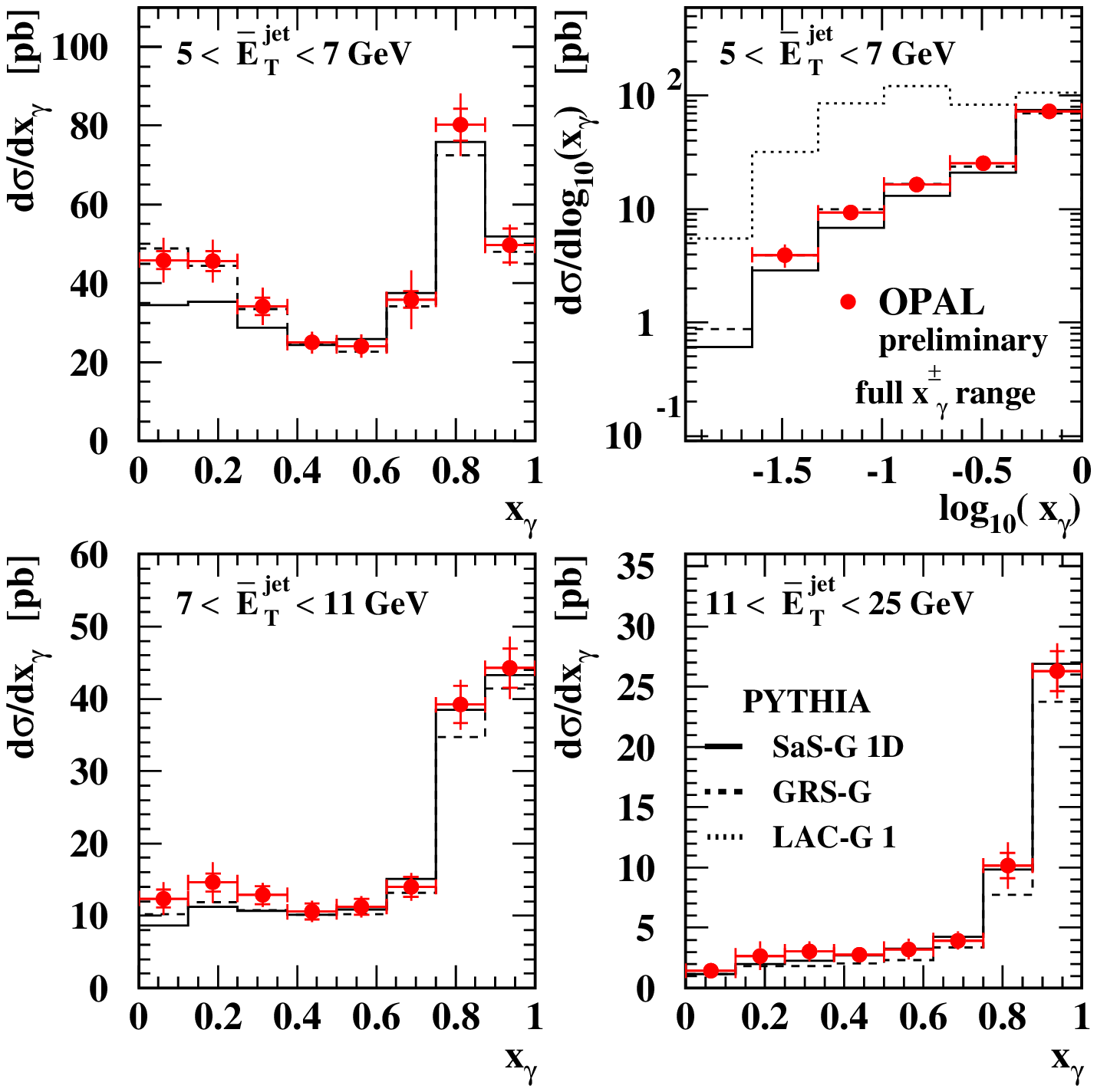} 
\epsfxsize=0.50\textwidth 
\epsfysize=0.58\textwidth 
\epsfbox{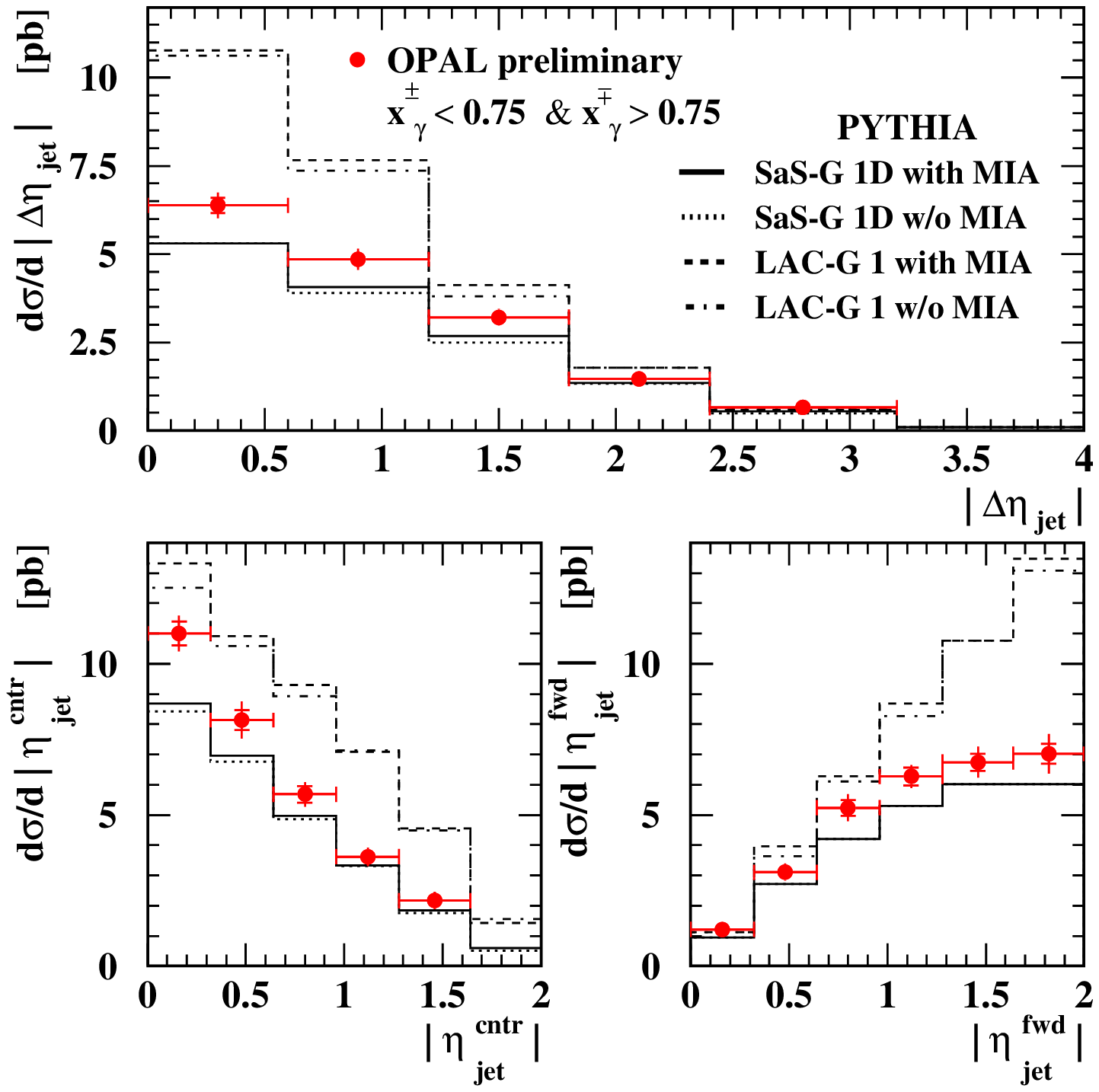} 
\caption{The dijet cross section (OPAL) as a function of $x_{\gamma}$ 
for the various mean transverse energies of the dijet system and 
as a function of $\Delta \eta$~\protect\cite{wengler}.
 \label{wengler-fig2}}
\end{figure}

Similar studies were also performed 
by DELPHI~\cite{masik}  for $E_T$ and jet profile distributions. 
Jet profiles turn out to be very sensitive to the presence of possible 
MIA effects. PYTHIA with MIA (default setting) 
and HERWIG with MIA (sue=20\%, i.e., in 20\% of  the simulated 
double-resolved events a soft underlying event was included) are favoured. 

 The OPAL and DELPHI data, for which comprehensive studies 
and Monte Carlo comparisons have been presented,  are 
consistent with each other.
Note however that comparisons to the NLO QCD calculation were made
 on parton level only~\cite{wengler} and  it had been pointed out previously  
 that hadronization corrections are important. The differential 
cross section in $E_t^{\mathrm {jet}}$ and $\eta$ need corrections of 
10\% to 20\%. However, for the distributions in $x_{\gamma}$, very sensitive to the hadronisation, the corrections may be even higher.\\

$\bullet${Jet and dijets production at HERA in photoproduction and DIS$_{\mathrm {ep}}$ 
events}\\

 At HERA using the jet data, parton densities  of the proton and 
 of the photon can be constrained. The structure of the real photon 
(photoproduction, $Q^2<1 \unit{GeV^2}$) and of the virtual 
photon (electroproduction or DIS$_{\mathrm {ep}}$ events with $Q^2> 1 \unit{GeV^2}$) 
have been studied, including direct and  resolved photon contributions. 
In the   resolved virtual photon processes at HERA the variable $Q^2$
is the virtuality (mass squared) of the photon, 
which structure is tested by large $p_T$ jets or particles, $p_T^2\gg Q^2$.
For the first time a 
partonic content of $\gamma^*_L$ was included in the analysis
of the HERA data, see below.
On the other hand interference terms, which may be important,
as it  was addressed in~\cite{jezuita}, were omitted in the analysis.

 H1 and ZEUS have measured cross section for the photoproduction of 
inclusive jet  as well as 
dijet   as a  function of $\eta$ and $E_T$~\cite{valcarova}. 
The ZEUS and H1 data are in nice agreement with each other in most 
observables  (after correction for different cuts) 
and with the calculations for inclusive jet production, see 
Fig.~\ref{valcarova-fig2b}-left and center, where H1 data are shown. 
However for dijets event   ZEUS finds a 
deficit at low $x_{\gamma}$ (even for large $E_T$ sample)
indicating that more gluons are needed in the photon.
It is  demonstrated 
in Fig.~\ref{valcarova-fig2b}-right, where the angular 
distribution is presented for various subprocesses. The H1 dijet 
data differ here from the ZEUS measurements and do not 
show a discrepancy with  the NLO predictions. 
\begin{figure}[t]
\epsfxsize=0.3\textwidth
\epsfysize=0.4\textwidth 
\epsfbox{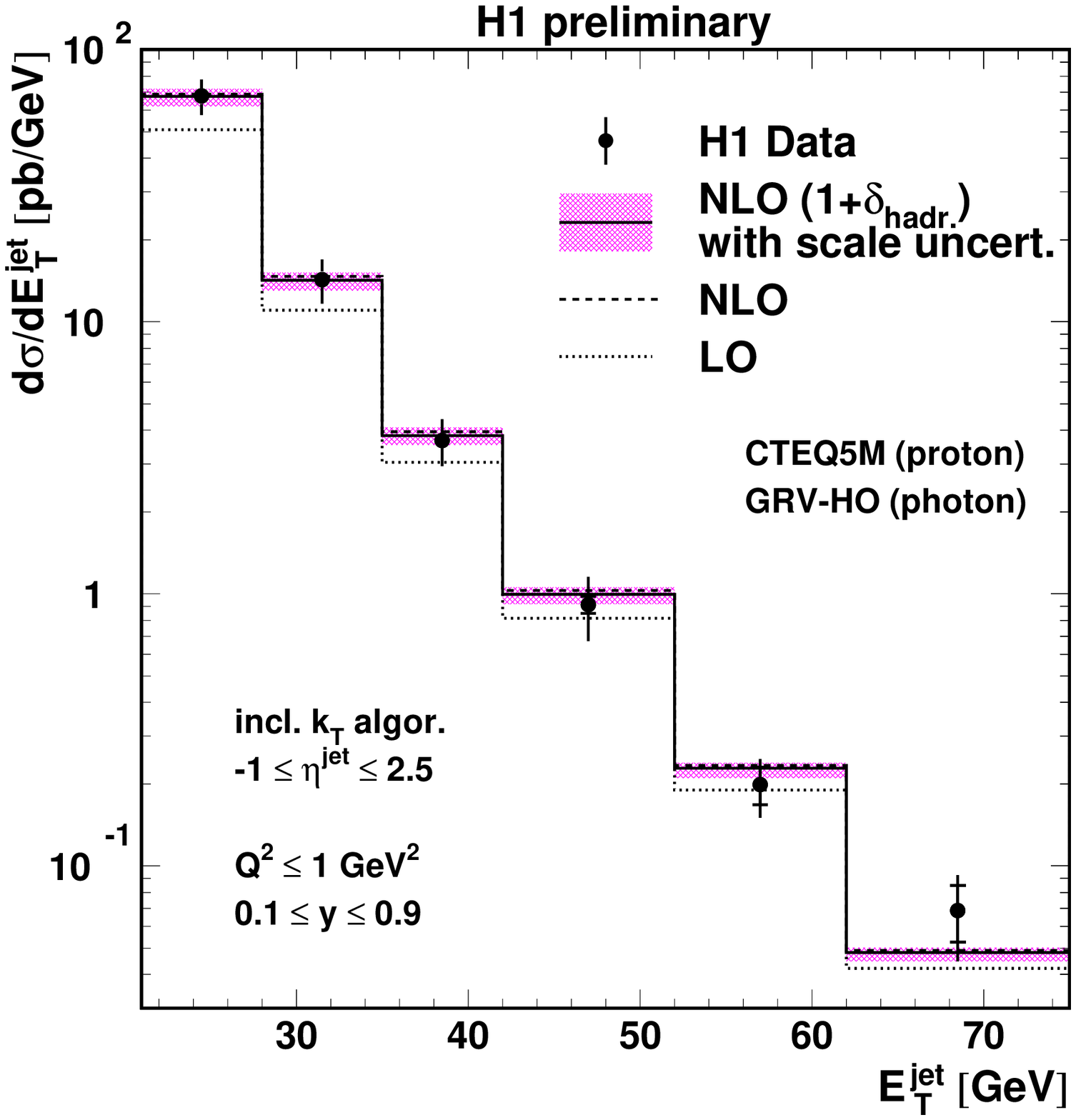}
\epsfxsize=0.3\textwidth
\epsfysize=0.4\textwidth 
\epsfbox{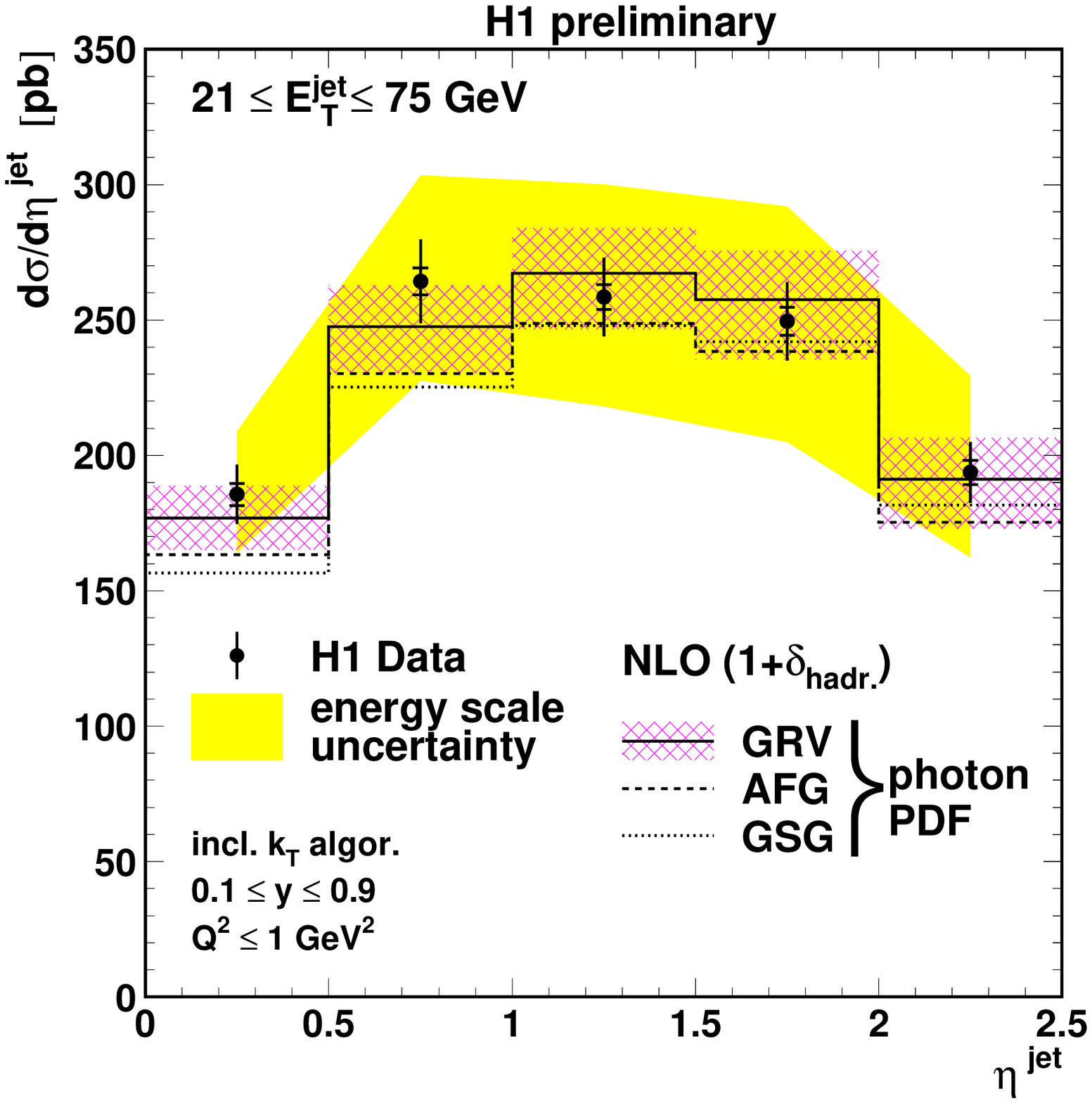}
\epsfxsize=0.3\textwidth
\epsfysize=0.4\textwidth 
\epsfbox{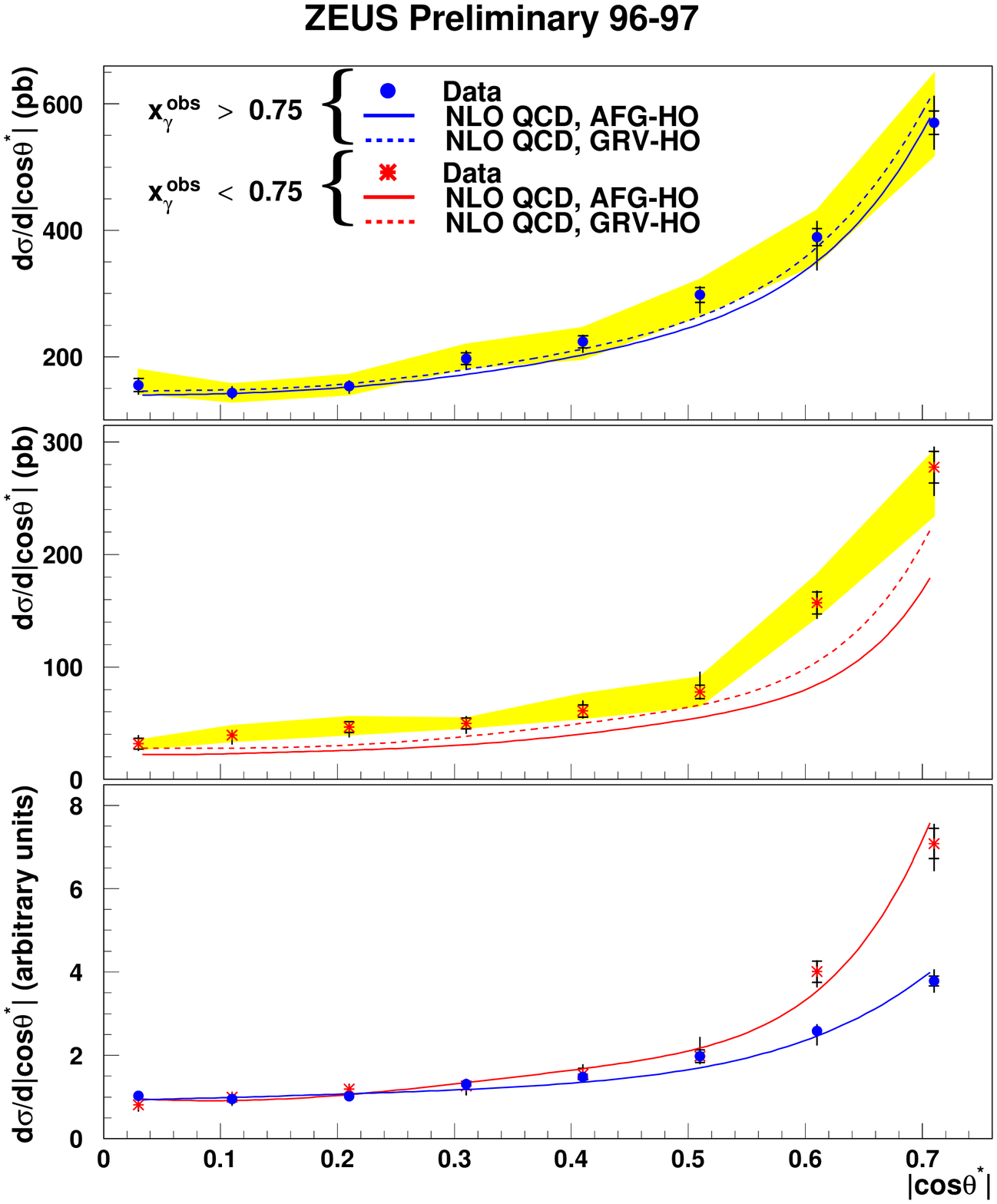}
\caption{Left and Centre: The  jet cross sections  for H1 data in comparison 
with the NLO calculations for  different parton 
parametrizations~\protect\cite{valcarova};\protect\\
Right: The dijet cross section (ZEUS) as a function of 
$|\cos\theta^*|$, from~\protect\cite{valcarova}. \label{valcarova-fig2b}}
\end{figure}
It was pointed out that 
the NLO calculations for cuts chosen for dijet events by ZEUS
(e.g., different $E_T^{min}$ for the two jets)
 should be more reliable~\cite{wing}.
\begin{figure}[t]
\begin{center}
\epsfxsize=0.49\textwidth
\epsfysize=0.60\textwidth 
\epsfbox{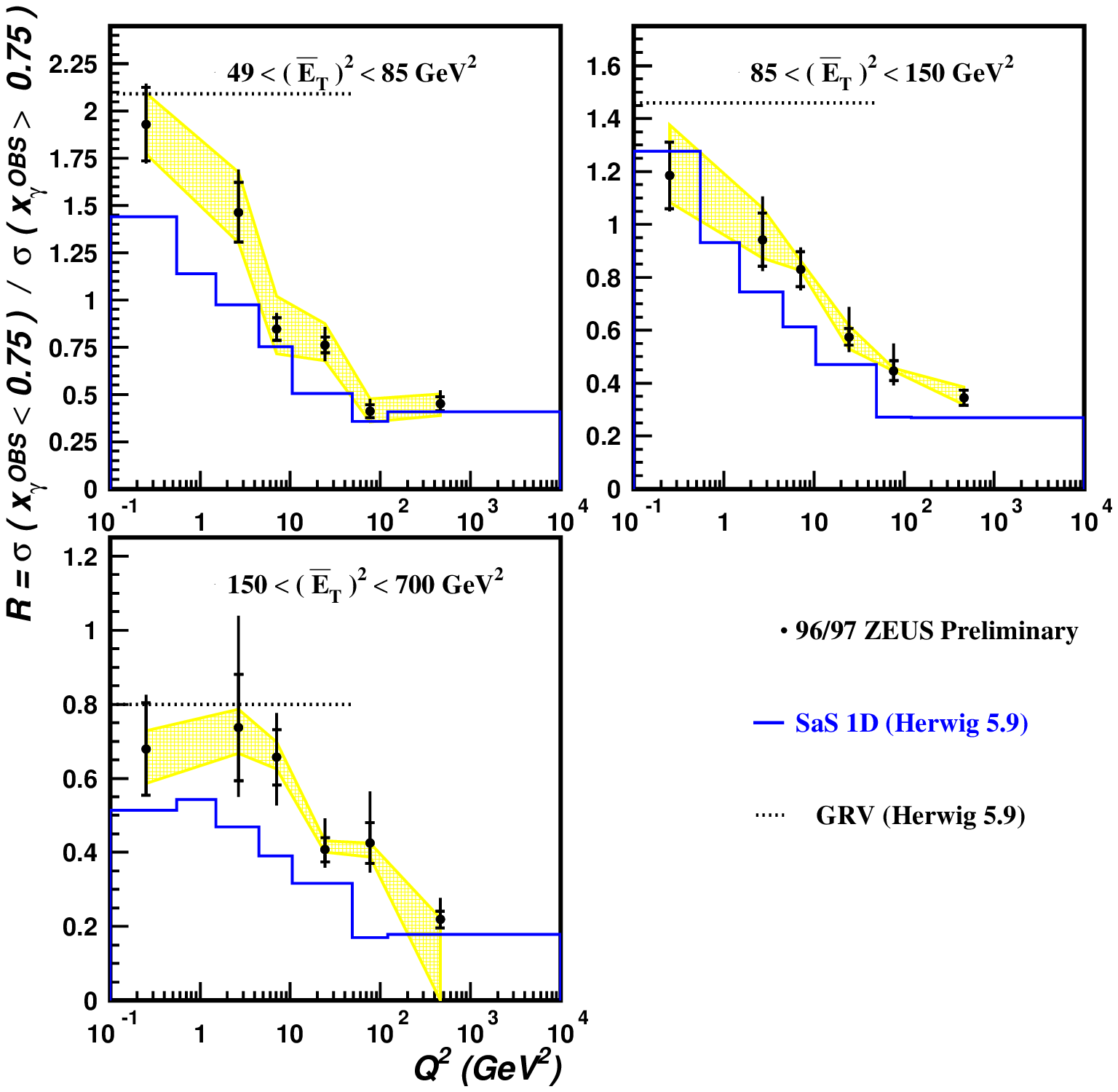}
\epsfxsize=0.49\textwidth
\epsfysize=0.60\textwidth 
\epsfbox{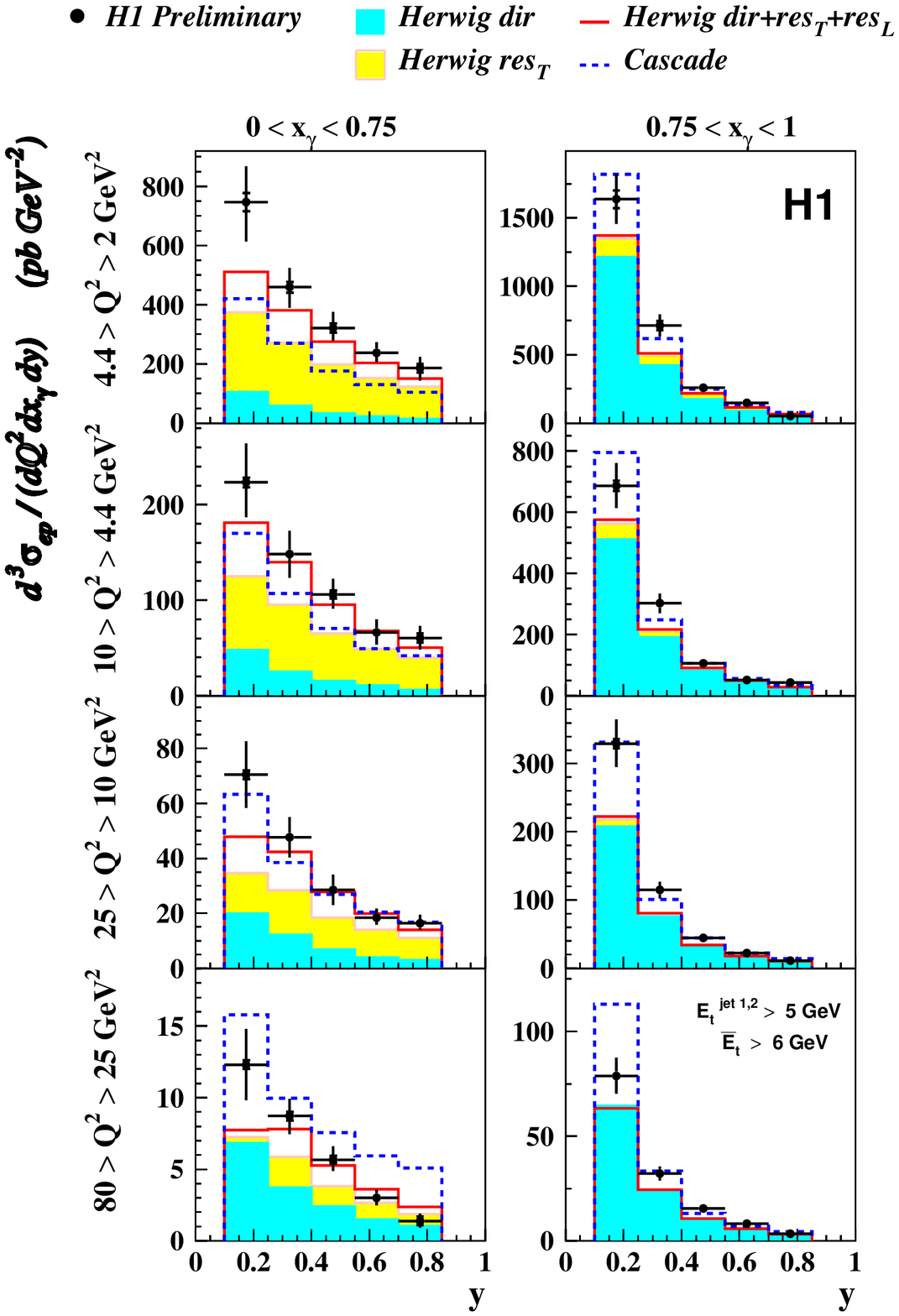}
\caption{
  Left: The ratio of cross sections
$R=\sigma(x_\gamma < 0.75) / \sigma(x_\gamma > 0.75)$, 
from ZEUS, as a function of $Q^2$~\protect\cite{sedlak};\protect\\
Right: Dijet cross section  for the H1 data,
the dark (light) histogram represents the direct $\gamma^*$ 
(direct plus the resolved $\gamma^*_T$)  contribution of HERWIG.
The full  line stands for the sum of the direct $\gamma^*$ and of $\gamma^*_T$
  and $\gamma^*_L$  resolved photon contributions of HERWIG;
 the dashed line is CASCADE prediction~\protect\cite{sedlak}.
}
\label{sedlak}
\end{center}
\end{figure}

The partonic content of the virtual photon 
has been investigated at HERA in dijet events,  confirming the
main $\ln p_T^2/Q^2$  dependence of the parton densities for $\gamma^*$  
predicted by the theory (already by Parton Model). This  behaviour leads  to   
a  suppression of the cross section with growing $Q^2$ for a fixed $p_T$ 
value (range).
ZEUS results, in form of a ratio of the cross sections for events with 
$x>0.75$ and
 $x<0.75$, are presented in Fig.~\ref{sedlak}-left (below similar ratio for  
subsample with charm will be shown). The ratio, plotted 
 as a function of $Q^2$ for various $E_T$ ranges,  
shows disagreement with HERWIG predictions 
(with SAS-1D, GRV). H1 data in two similar samples 
were compared with HERWIG MC predictions based on the   direct $\gamma^*$
and  on the resolved $\gamma^*_T$ contribution.
When in addition the contribution due to the  partonic content  of 
the longitudinally polarized virtual photon is included in 
analysis~\cite{sedlak}, the  description of the data is improved. 
Here also the CASCADE Monte Carlo  based on  CCFM evolution  was applied
- it gives prediction even 
closer to the data (Fig.~\ref{sedlak}-right),
however the $Q^2$ dependence is not well reproduced~\cite{sedlak}.

\subsection{Prompt Photons}

\begin{figure}[t]
\epsfxsize=0.6\textwidth
\epsfysize=0.45\textwidth 
\epsfbox{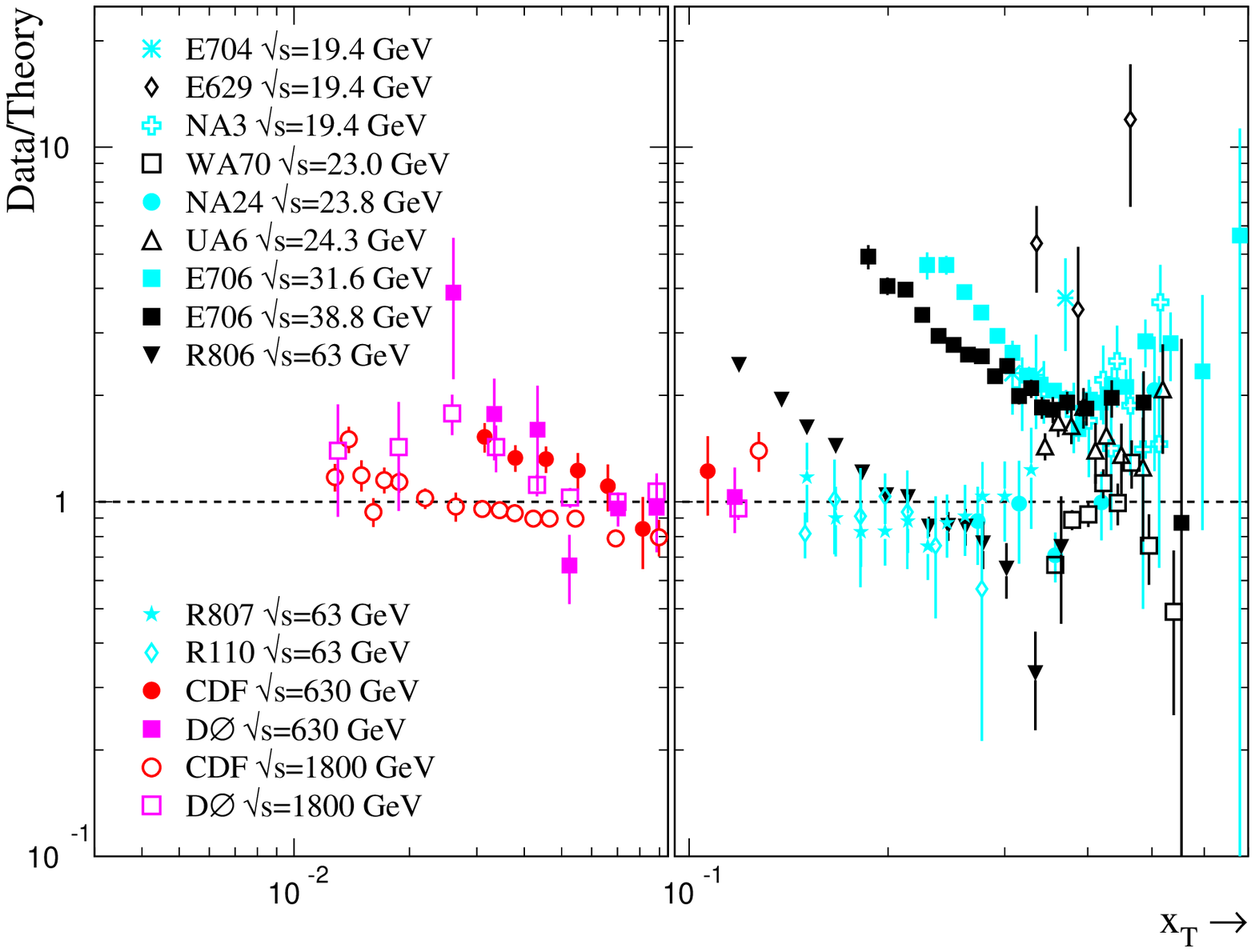}
\epsfxsize=0.45\textwidth
\epsfysize=0.5\textwidth 
\epsfbox{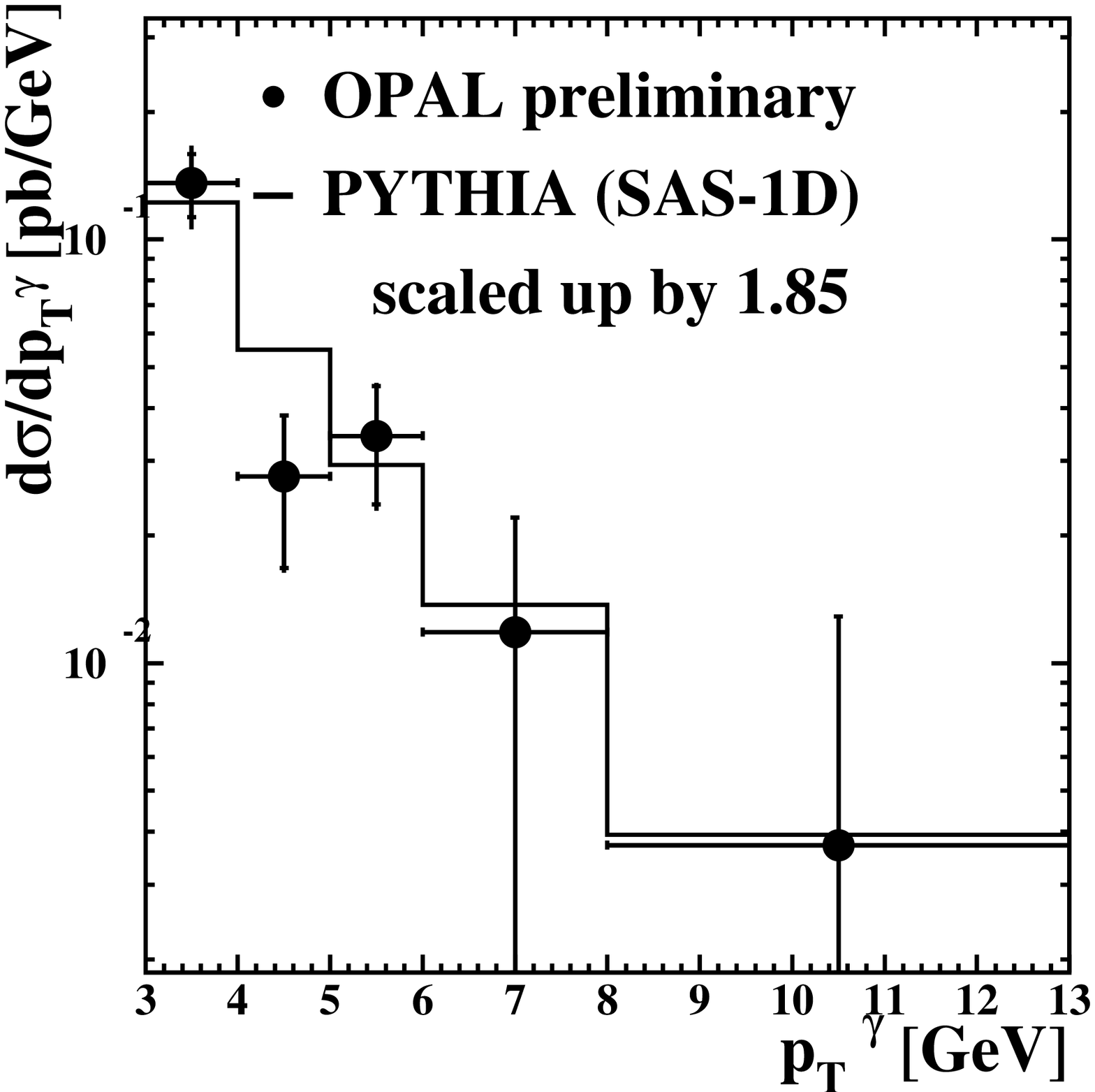}
\caption{ The prompt photon production. Left: At hadron-hadron 
colliders in comparison with NLO QCD prediction as function of 
$x_T = 2E_T/\sqrt{s}$, from~\protect\cite{begel};\protect\\
Right: At LEP as a function of 
$p_T$ in comparison with PYTHIA results~\protect\cite{lillich}. 
\label{begel-fig}}
\end{figure}

At  the Tevatron, HERA  and recently also at LEP collider the prompt 
(isolated) photon production has been studied~\cite{begel,lillich}.
While for  ${\mathrm {p\bar p}}$ collisions the   
D0 data agree with the NLO predictions, 
CDF observes an excess at low photon $p_T$, 
similar as in  measurements 
at lower centre-of-mass energy in hadron-hadron 
collisions, see  Fig.~\ref{begel-fig}-left. 
The $k_t$ smearing for initial partons helps to describe data. 
The run~II at Tevatron may 
clarify the situation, see also~\cite{tapprogge} where plans of future 
measurements of prompt photons at LHC are discussed. Note, that  such processes 
are sensitive to the gluonic content of the proton.
 At the ep-collider HERA the photoproduction of the prompt photon 
plus jet  were measured by the ZEUS group recently 
(in previous publications ZEUS~\cite{zeus-p} dealt with prompt photons only; then, their 
data  showed 
 slight disagreement with the NLO calculation in the $\eta$ distribution 
for small $y$ sample~\cite{f-z}).
In this new measurement  a direct 
photon sample, ie. with $x_{\gamma}\sim 1$, was used to extract
  the  $\langle k_t \rangle , \sim 2 \unit{GeV}$, in the proton~\cite{zeus-k}. 
The new 
OPAL measurements of the inclusive production of the isolated photon in the process $\gamma \gamma \rightarrow \gamma$X 
needs a $K$-factor of 1.85  to describe the data 
(Fig.~\ref{begel-fig}-right)~\cite{lillich}.

\subsection{ Hadron  Production}

\begin{figure}[t]
\epsfxsize=0.45\textwidth
\epsfysize=0.4\textwidth 
\epsfbox{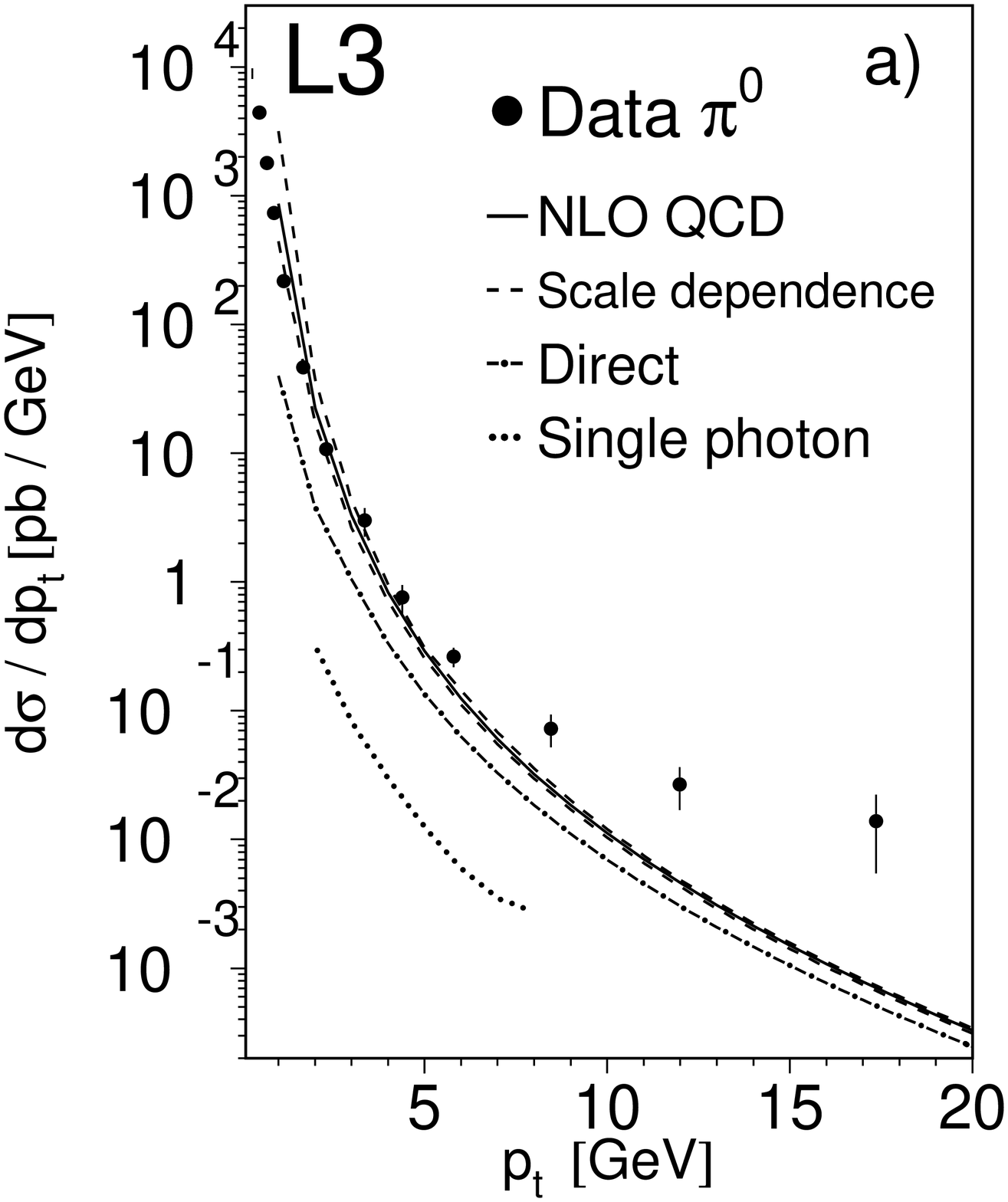} 
\epsfxsize=0.45\textwidth 
\epsfysize=0.4\textwidth 
\epsfbox{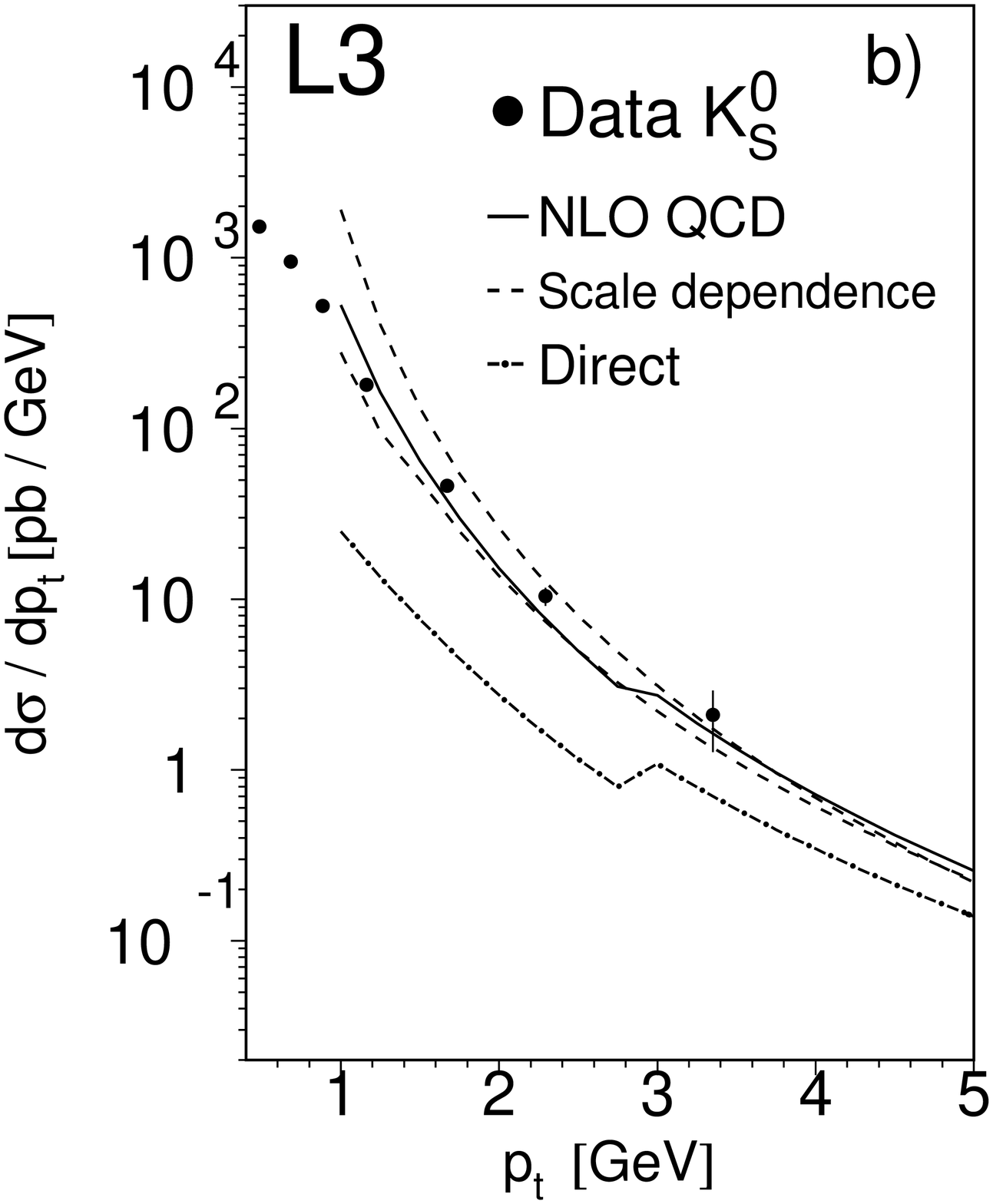} 
\caption{Inclusive differential cross section ${\mathrm d}\sigma/
{\mathrm d}p_T$ for (a) $\pi^0$  and (b) K$_{\mathrm S}$  in comparison with NLO QCD 
calculations~\protect\cite{achard}. \label{achard-fig1}}
\end{figure}

In Fig.~\ref{achard-fig1} $p_T$ spectra for $\pi^0$ and K$_{\mathrm S}$ 
in two-photon collisions are presented as measured by 
L3. The data  show 
the expected exponential behaviour for small $p_T$,
for $p_T > 1.5\unit{GeV}$ a power law $p_T^{-B}$ 
with $B \approx 4$ indicates  a simple underlying
 2 body $ \rightarrow 2 $ body process. The 
spectra are also compatible with NLO calculation  or PHOJET 
prediction (which are similar) 
for momenta up to a few GeV. Beyond $5\unit{GeV}$, where 
 data only for $\pi^0$ are available 
a clear disagreement is seen:
the NLO prediction lies below the data points (Fig.~\ref{achard-fig1}-a),
PHOJET exceeds the measurements while PYTHIA is below 
(not shown)~\cite{achard}. 
It was pointed out that further tests for charged pions are welcome.

\section{Charm and Beauty Production}
Heavy quark production provides important tests of QCD. It gives 
complementary to jets information on partonic densities in the photon.
Although the large quark mass should allow for
 reliable perturbative calculations in QCD, discrepancies are observed 
for the heavy quark productions, 
especially for b production, at all colliders.
The question may appear on a
limitation of  the standard calculation  based on a DGLAP evolution,
both for massless and massive approaches  used in the calculations.
It has been pointed out at this conference that a recently 
developed 
CASCADE program based on  the 
CCFM evolution equations using unintegrated parton distributions 
reproduces the b production cross section at 
the Tevatron~\cite{jung}. 
CASCADE also reproduces most charm and bottom distributions at HERA,
unfortunately it has not been applied for $\gamma \gamma$ processes. 

For beauty production the QCD predictions are expected to be more reliable 
than for charm, but are still limited by the smaller cross section.
More considerations for a better description of  in the b sector were debated
(e.g., possible need of additional terms~\cite{chyla}).
It was emphasized~\cite{frixione} that care should be taken, whether comparisons are 
performed for B meson spectra or b-jet spectra, with the latter being 
more closely related to the partons. 

A clear, non-ambiguous signal for charm quark production is the presence of
a ${\mathrm{D}}^{*+}$. 
This gold-plated signature has been exploited
by all four LEP and both HERA collaborations.
The leptonic decay has been used by ALEPH ($\mu^{\pm}$ for charm), L3
($\mu^{\pm}$ and ${\mathrm{e}}^{\pm}$ both for charm and bottom) and 
OPAL ($\mu^{\pm}$for bottom). Though
these analyses rely on the momentum spectrum as theoretical input, the
statistics is substantially increased as compared to the ${\mathrm{D}}^{*+}$ 
analyses~\cite{andreev,sokolov}.
The b measurements at HERA rely on the inclusive semi-leptonic decays, 
using electrons or muons in dijet events. This signature 
is combined either with the requirement of a large 
$p_{\mathrm T}^{\mathrm{rel}}$ to a jet (exploiting the effect of 
the large b quark mass), 
or the secondary vertices (profiting from the lifetime of the B mesons).

\begin{figure}[t]
\epsfxsize=0.3\textwidth
\epsfysize=0.45\textwidth 
\epsfbox{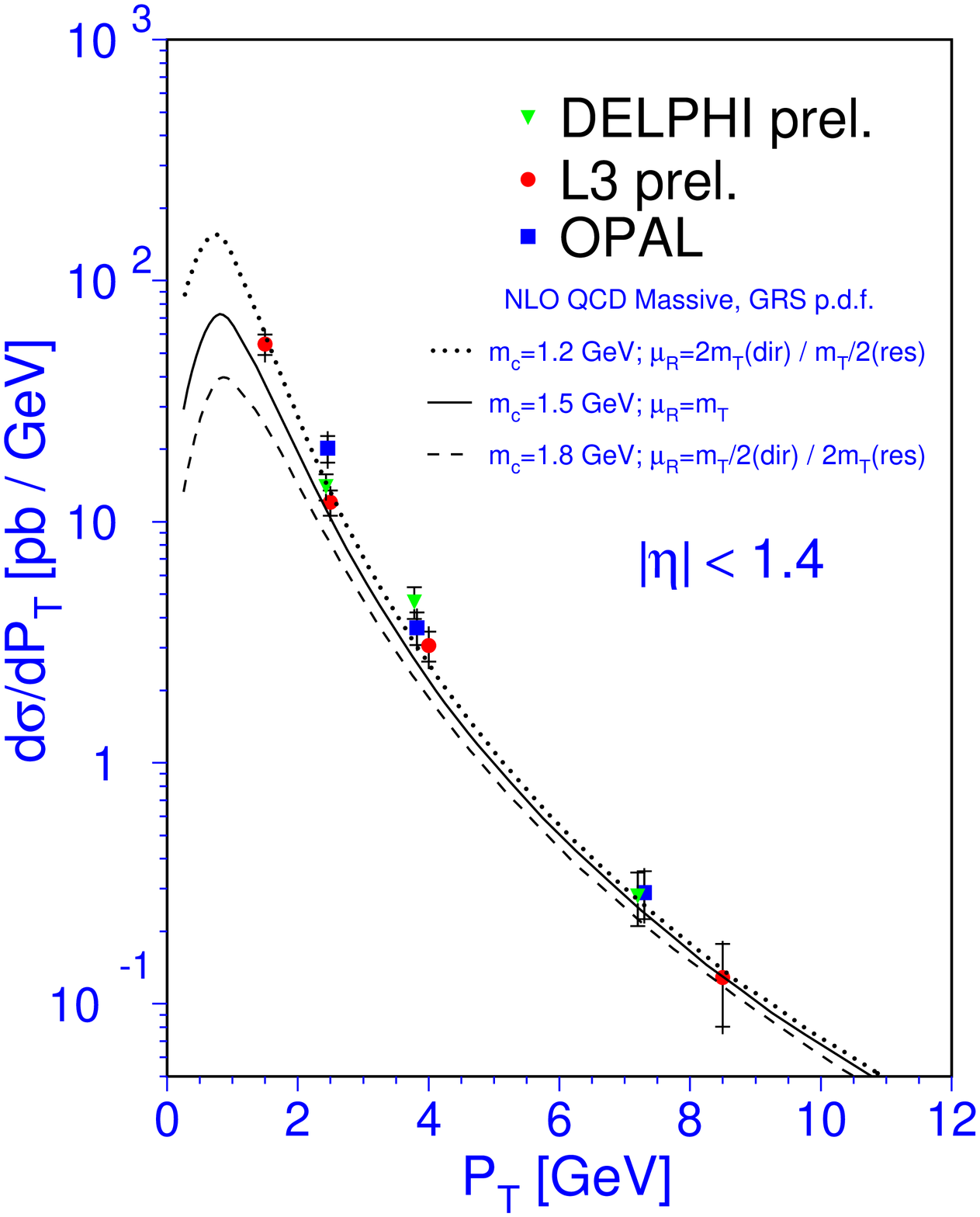}
\epsfxsize=.3\textwidth
\epsfysize=.50\textwidth
\epsfbox{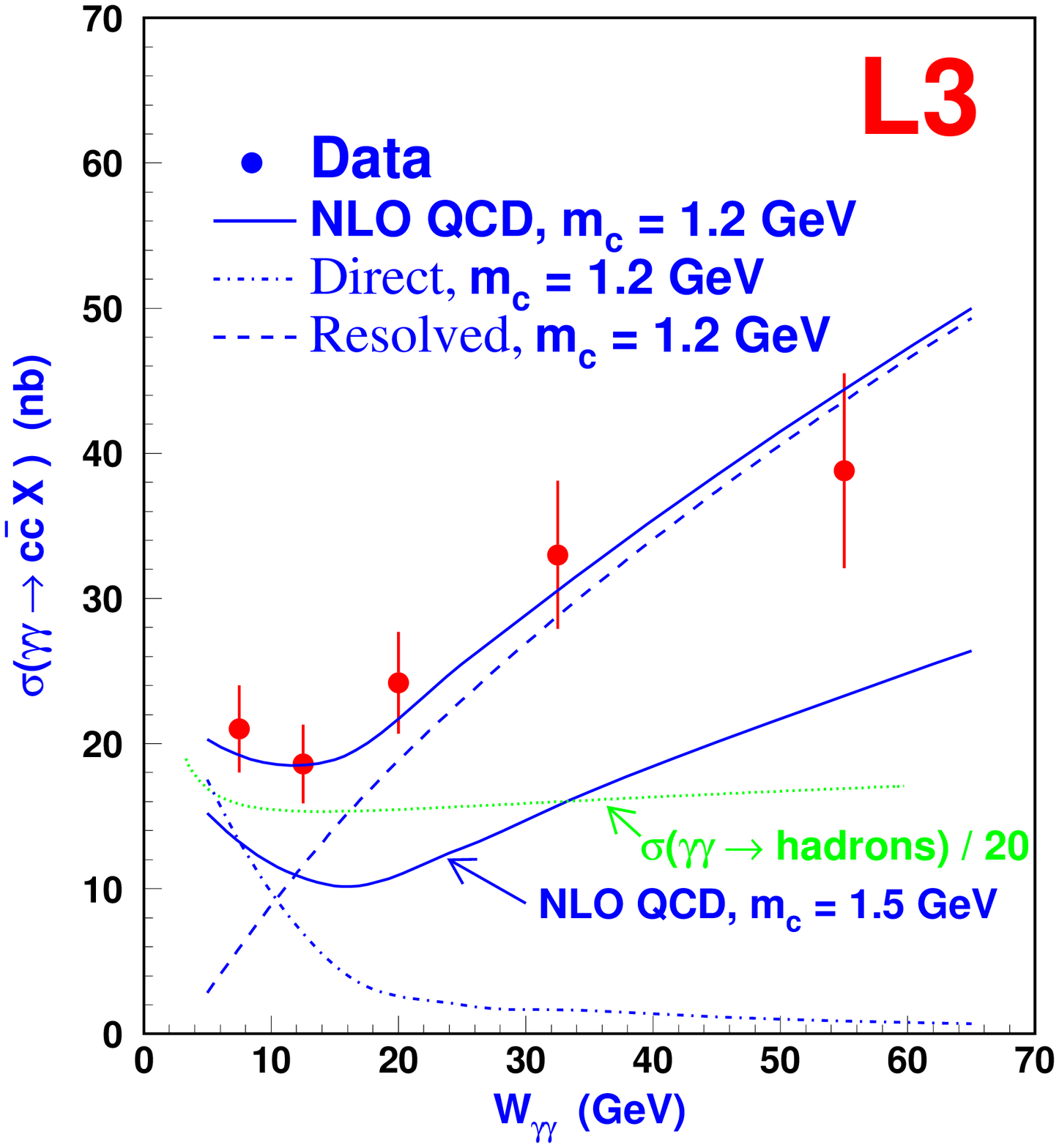}
\epsfxsize=0.3\textwidth
\epsfysize=0.50\textwidth 
\epsfbox{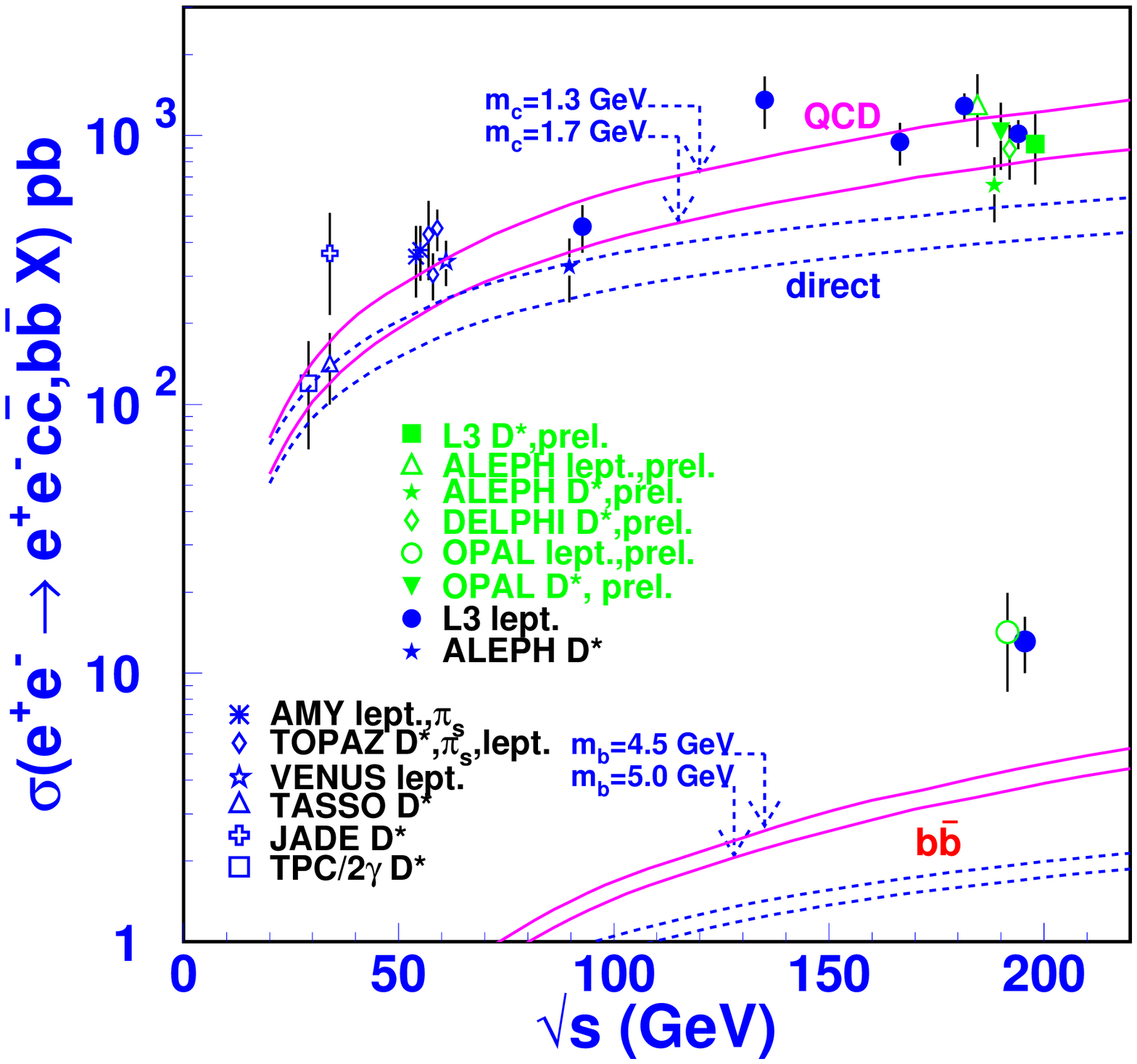}
\caption{Left: Differential cross section for D$^*$ production obtained 
by LEP experiments in comparison with NLO QCD calculations, Centre:
 Charm production in $\gamma \gamma$ collision;\protect\\
Right: Summary of open charm and bottom production at e$^+$e$^-$ 
colliders in comparison with NLO QCD 
calculations~\protect\cite{andreev}.
 \label{andreev-fig3}}
\end{figure}

\subsection{ Heavy Flavour Production at LEP}

As discussed above for jets the production of heavy flavour in 
two-photon collisions is dominated
by two processes, the direct and the single-resolved process. 
Both contribute
in equal shares to heavy flavour final states at
LEP~2 energies. 
The LEP experiments, ALEPH, DELPHI, L3, and OPAL,
agree among themselves and with the theoretical expectation, which e.g.,
 predicts 
a flat distribution in pseudorapidity for D$^*$ production.
In transverse momentum distribution 
the measurements are close to the NLO QCD prediction based on a massive quarks 
approach, while there is some scatter among data,
Fig.~\ref{andreev-fig3}-right.
The experiments measure the fraction of direct and single-resolved 
contribution on a 10\% level and confirm the prediction of NLO 
calculations~\cite{frixione}.

The charm cross section as function of $W_{\gamma\gamma}$
(Fig.~\ref{andreev-fig3}-centre) shows a steeper 
rise as compared to $\sigma (\gamma\gamma \rightarrow$hadrons), see Sec.~5.
 With 
$m_{\mathrm {charm}} = 1.2\unit{GeV}$ the agreement with NLO QCD calculation
is very good and 
also shows that the contribution from the resolved process dominates at high 
$W_{\gamma\gamma}$~\cite{kienzle}.

The present status of the measurements for the total 
${\mathrm {e^+e^-\rightarrow e^+e^- c \bar c, b\bar b X}}$ 
cross section is summarized in 
Fig.~\ref{andreev-fig3}-right. The importance of the resolved
photon contribution is obvious for both charm and bottom production.
The results are compared to NLO calculations, and good agreement is seen 
for charm production while a large discrepancy (3-4 $\sigma$) is observed 
for bottom. 
Bottom production is measured by  L3  using the
fact, that the momentum as well as the transverse momentum of leptons
with respect to the closest jet is higher for muons and electrons from bottom
than for background, which is mainly charm. The total bottom production 
cross section is extracted correcting for detector efficiency. 
Note a large sensitivity of the cross section to the value of heavy quark mass.
The same findings on agreement with theory, 
 but with the 
studies restricted to the muonic channel, come from OPAL. This experiment 
also investigated the $x_T = 2p_T/W_{\mathrm {vis}}$ distribution, where 
$p_T$ is the transverse momentum of the muon with respect to the beam axis; 
the shape of the MC simulation
is found compatible with the data~\cite{andreev}.

\subsection{Heavy Flavour Production at HERA}

\begin{figure}[t]
\epsfxsize=0.35\textwidth
\epsfysize=0.58\textwidth
\epsfbox{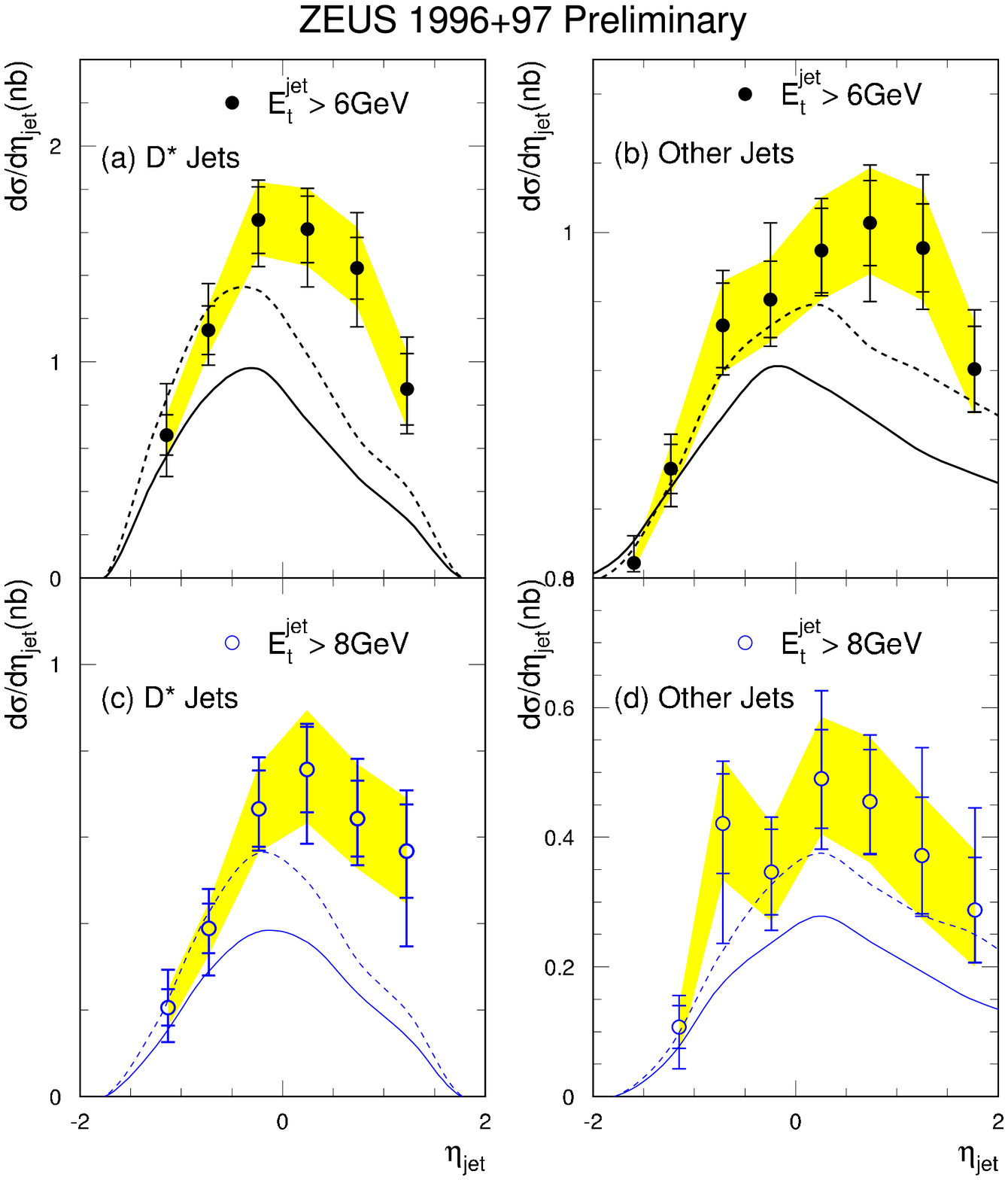}
\epsfxsize=0.25\textwidth
\epsfysize=0.58\textwidth 
\epsfbox{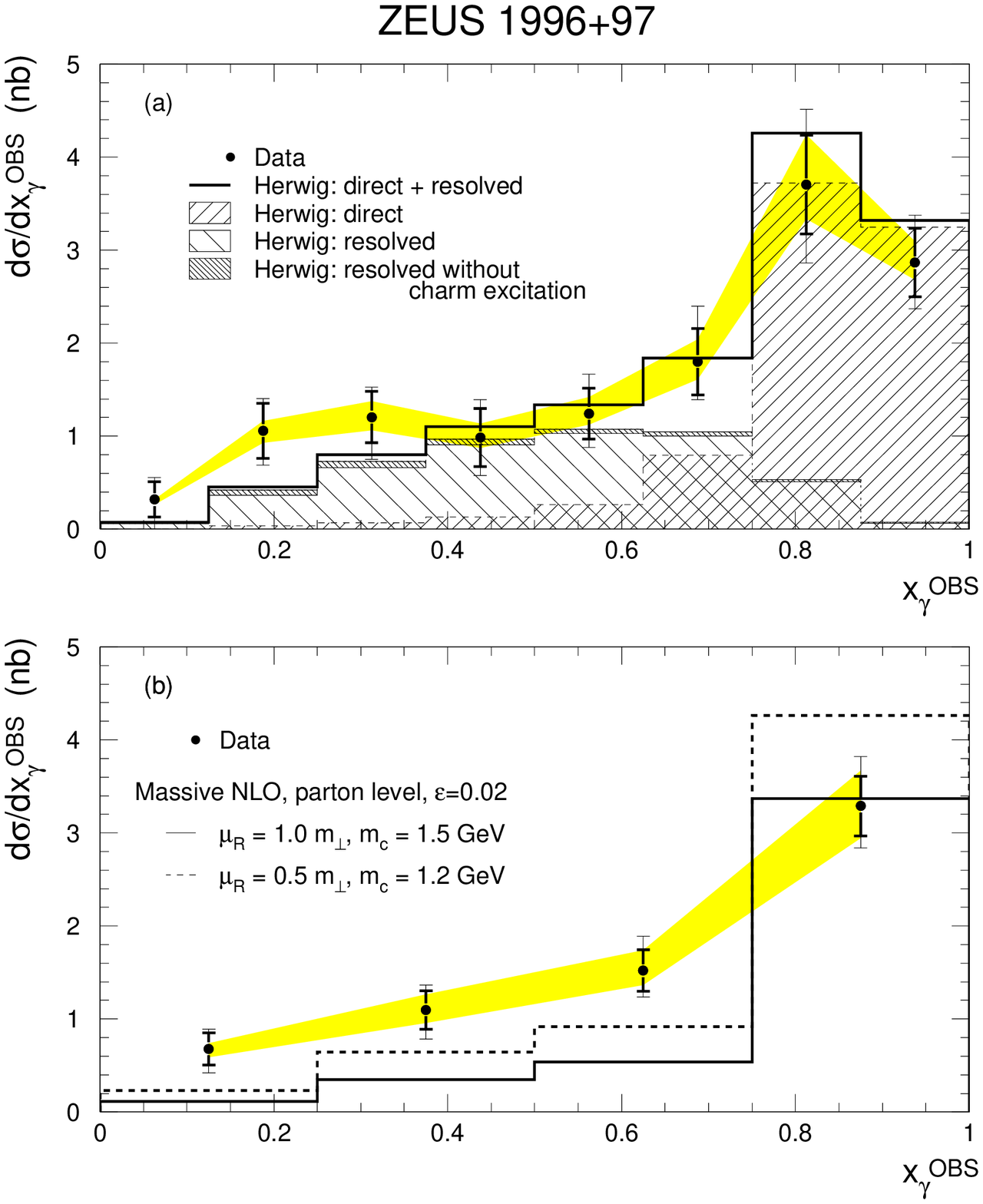} 
\epsfxsize=0.30\textwidth 
\epsfysize=0.58\textwidth 
\epsfbox{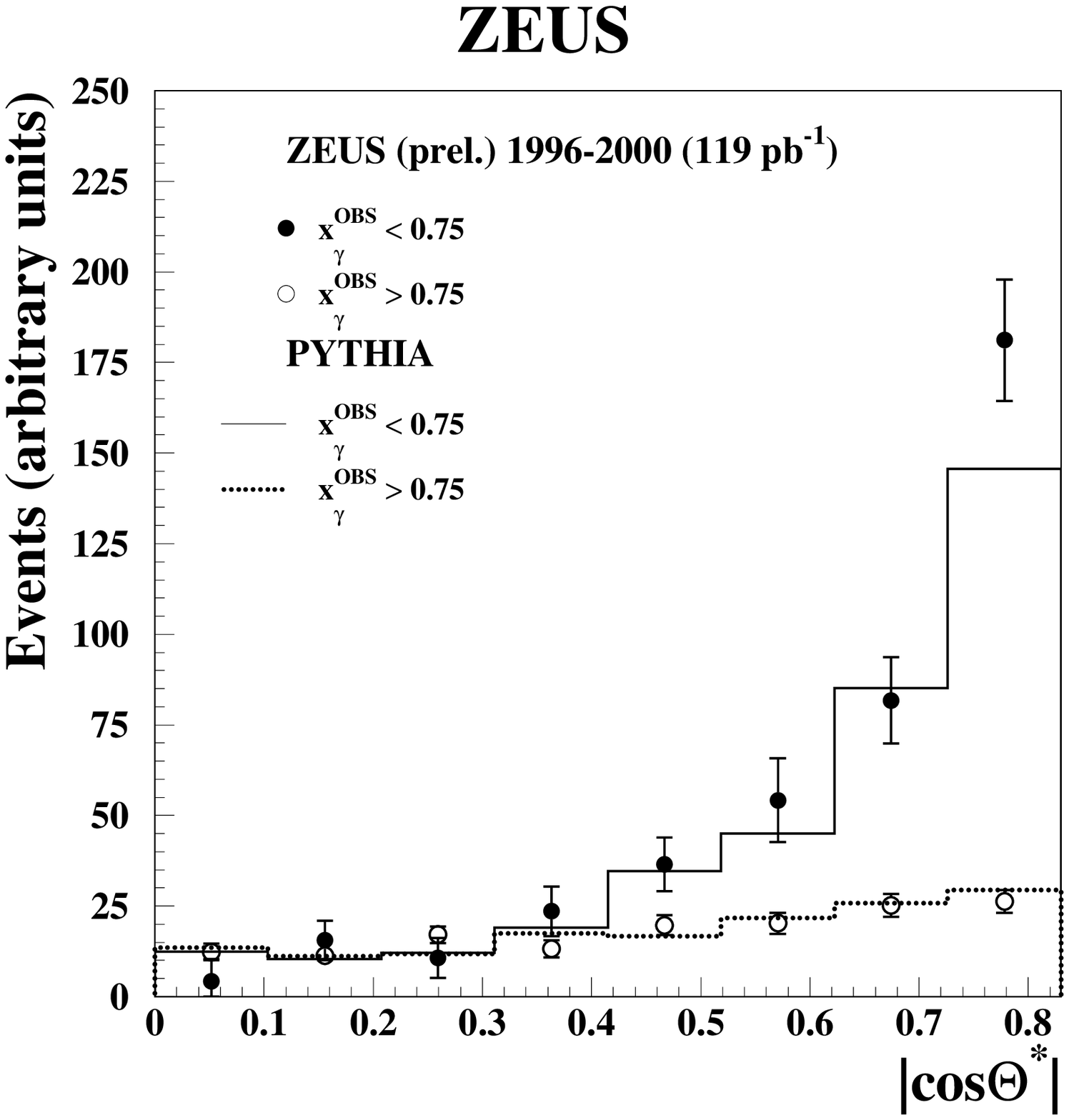} 
\caption{Charm in photoproduction (ZEUS). Left: Data for charm and non-charm jets;\protect\\
Centre: Results for dijets 
for with a reconstructed D$^*$ meson in comparison with 
the HERWIG model for $x_{\gamma}$ distribution
 (a) and with a parton level NLO QCD calculation (b);\protect\\
Right: 
The $\cos \theta^*$ distributions  for dijets events with additional cuts;
direct and resolved $\gamma$ samples in comparison with the PYTHIA model~\protect\cite{padhi}.\label{padhi-dijet}}
\end{figure}

The charm production  (jets and D$^*$ mesons) 
are measured by H1 and ZEUS both for a real and 
virtual photon interaction with a proton. 
Although such processes are dominated by the photon-gluon 
fusion, which  directly probes the gluon content of the proton 
(or the pomeron for the diffractive events) they also 
 offer opportunity to study the resolved photon contributions.  

\begin{figure}[t]
\epsfxsize=0.35\textwidth
\epsfysize=0.5\textwidth
\epsfbox{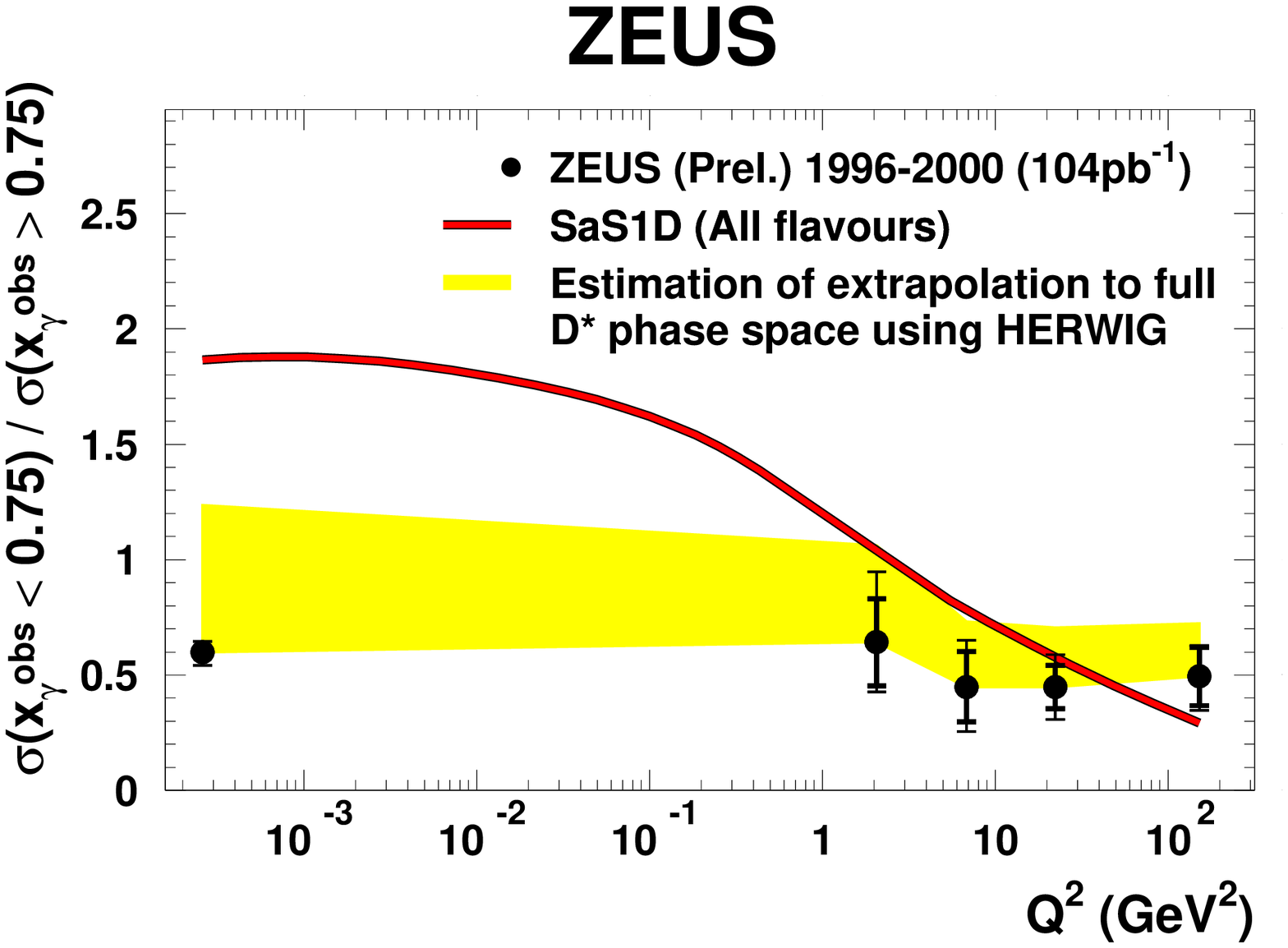}
\epsfxsize=0.3\textwidth
\epsfysize=0.45\textwidth
\epsfbox{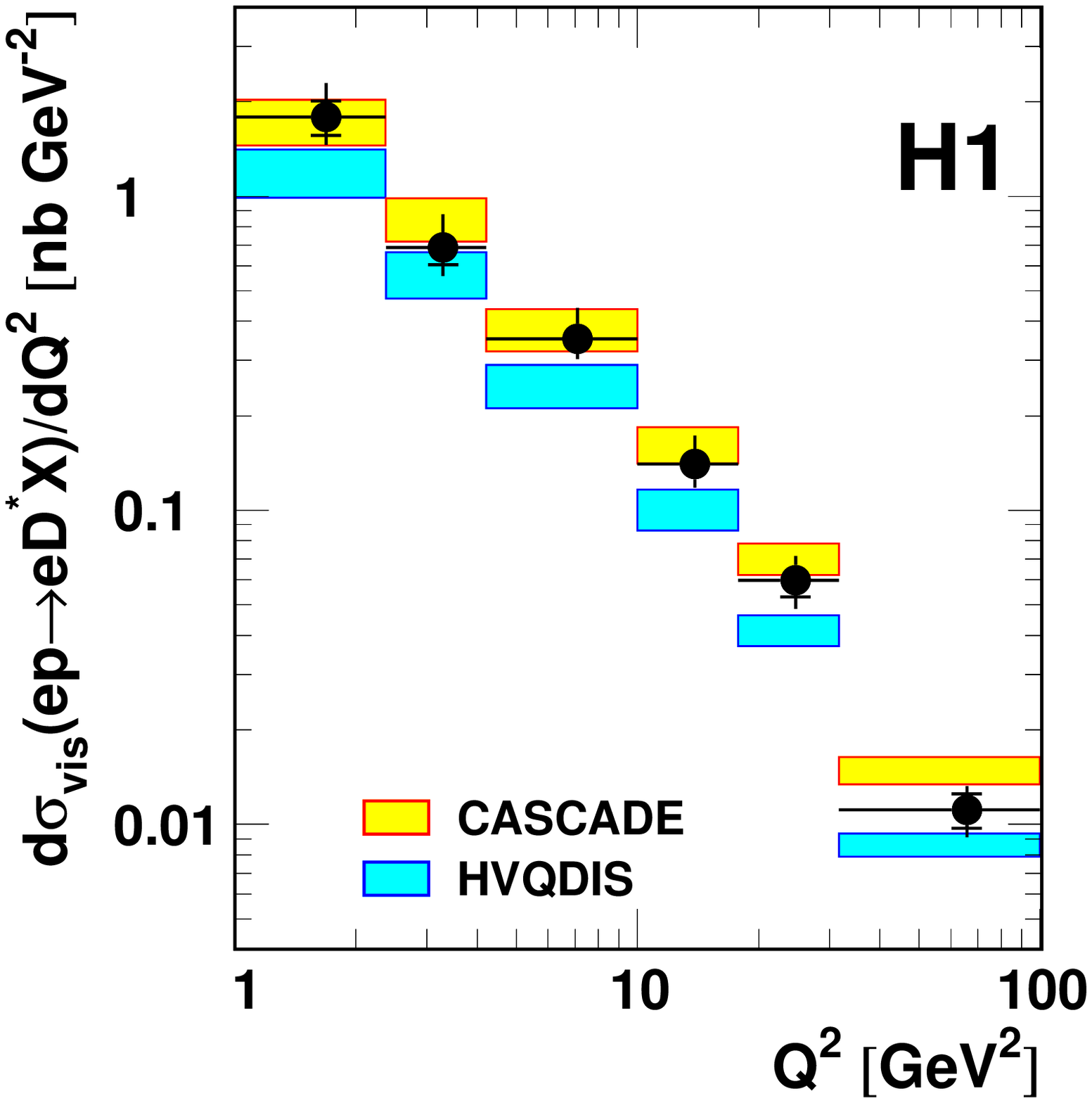}
\epsfxsize=0.3\textwidth
\epsfysize=0.45\textwidth
\epsfbox{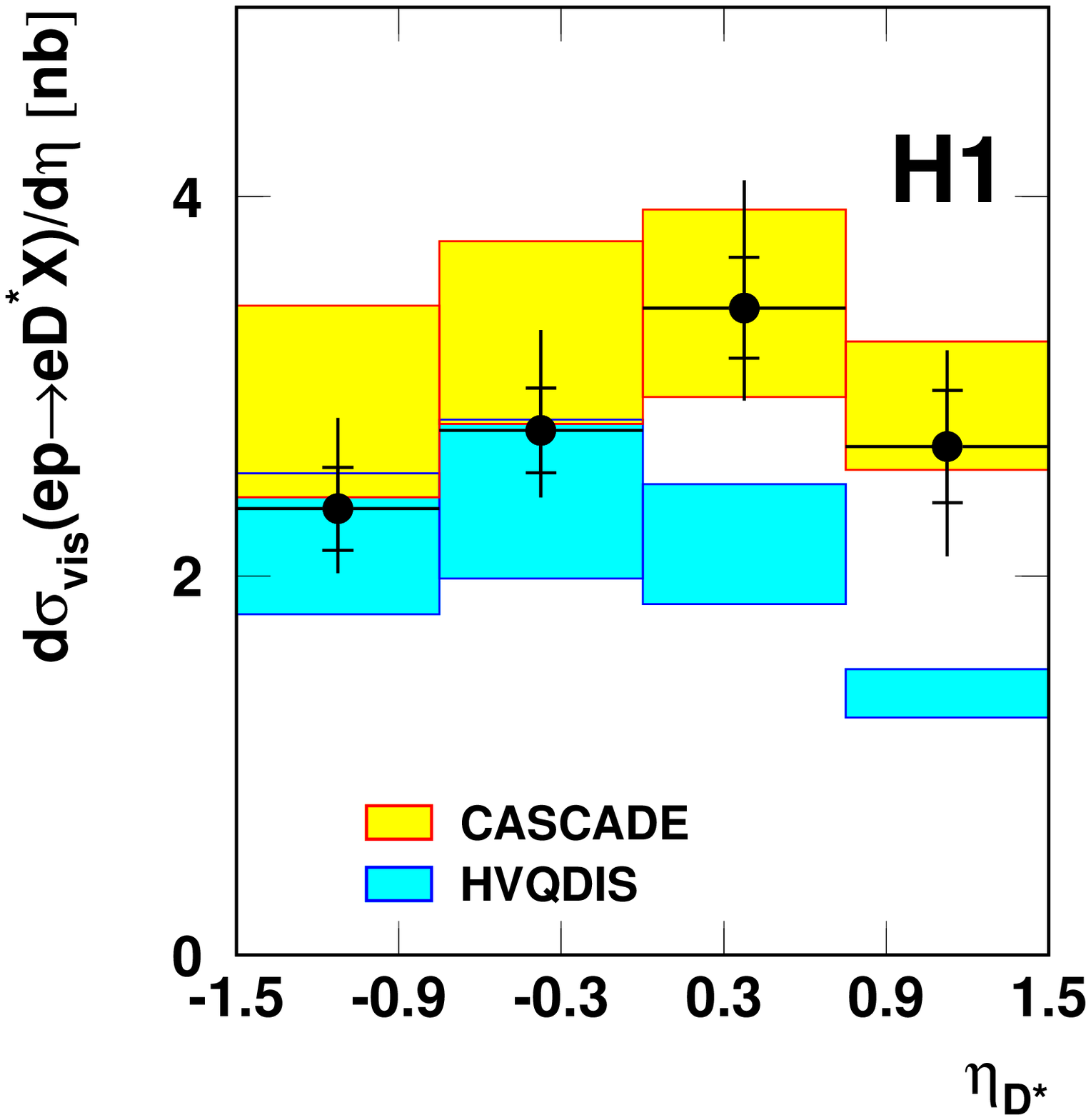}
\caption{Charm in DIS$_{\mathrm {ep}}$. Left: Ratio of 
dijet events with a reconstructed D$^*$ 
compared to  predictions of the SaS-1D parton 
parametrization for $\gamma$~\protect\cite{west};\protect\\
Centre: Results for D$^*$ production as a function of $Q^2$;\protect\\
Right: as in center for $\eta$, from~\protect\cite{erdmann}. 
\label{west-fig2}}
\end{figure}

 The  cross section
as measured by ZEUS in photoproduction as a function of pseudorapidity for charm and non-charm jets are shown in
Fig.~\ref{padhi-dijet}-left. The failure of the NLO QCD  to describe
the data in the forward hemisphere is evident. The difference of shape 
and normalization can not
be explained by the fragmentation, not included in the
simulation~\cite{padhi}. 
A similar effect is seen in Fig.~\ref{padhi-dijet}-centre; 
HERWIG and NLO QCD fail to describe the data
for low $x_{\gamma}$, where the resolved photon contribution dominates. 
Nevertheless, checking the underlying dynamics by the
$\cos \theta^*$ distribution,  with additional cuts (e.g., $M_{JJ} >$ 18 GeV), 
 separately for direct (large $x_{\gamma}$) and
resolved (low $x_{\gamma}$; gluon exchange) enriched samples,
good agreement with the PYTHIA simulation is found, see 
Fig.~\ref{padhi-dijet}-right.

For the virtual photon the ratio of low to high $x_{\gamma}$ in 
dijet events as function of $Q^2$ (Fig.~\ref{west-fig2}-left)
was measured by ZEUS, and 
suppression with $Q^2$ was found,
 weaker however than for the all-flavours case~\cite{west}. 
Suppression with growing $Q^2$ is observed also in H1 data for 
D$^*$ production, see Fig.~\ref{west-fig2}-centre.

\begin{figure}[t]
\epsfxsize=0.50\textwidth
\epsfysize=0.50\textwidth 
\epsfbox{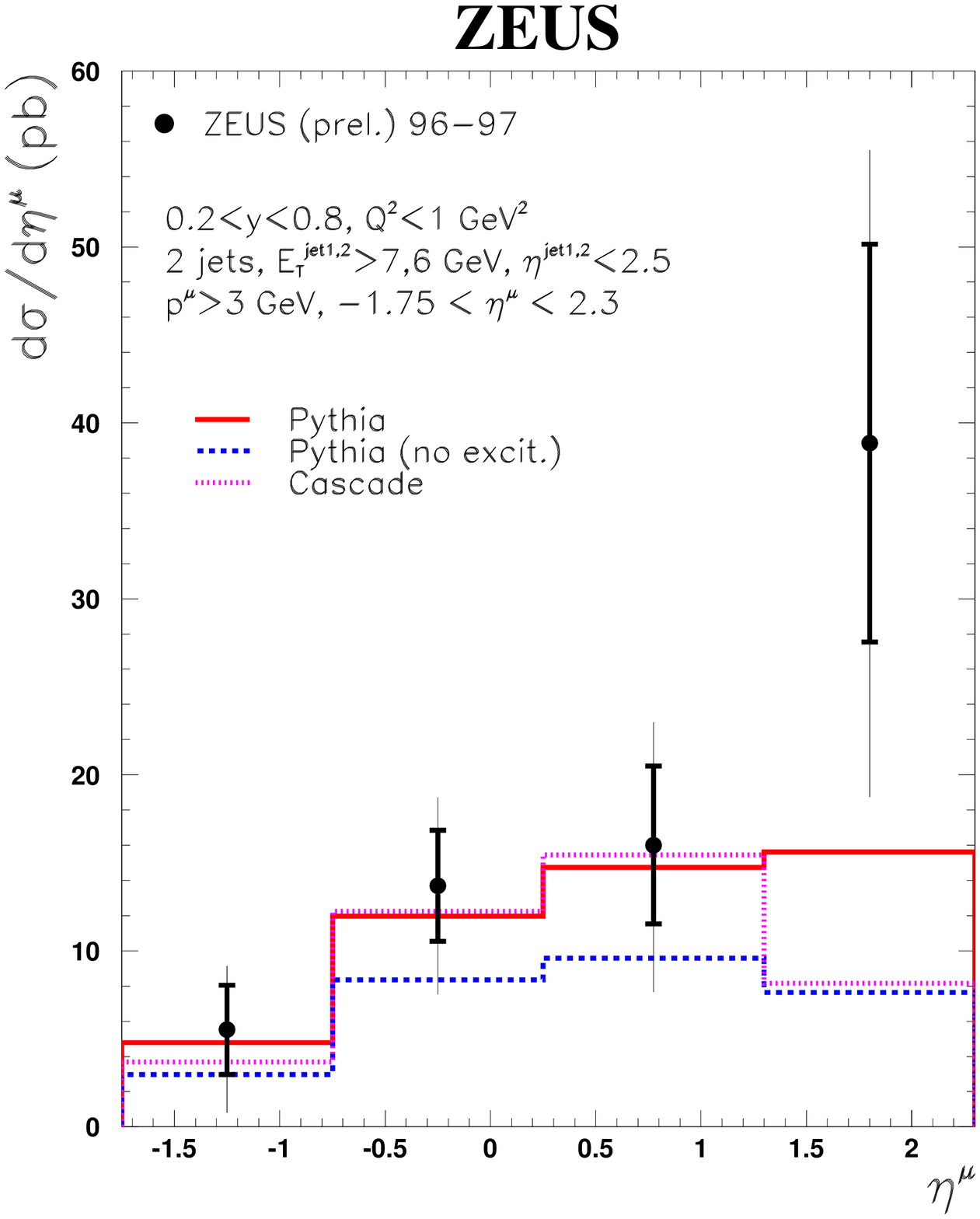}
\epsfxsize=0.50\textwidth
\epsfysize=0.50\textwidth
\epsfbox{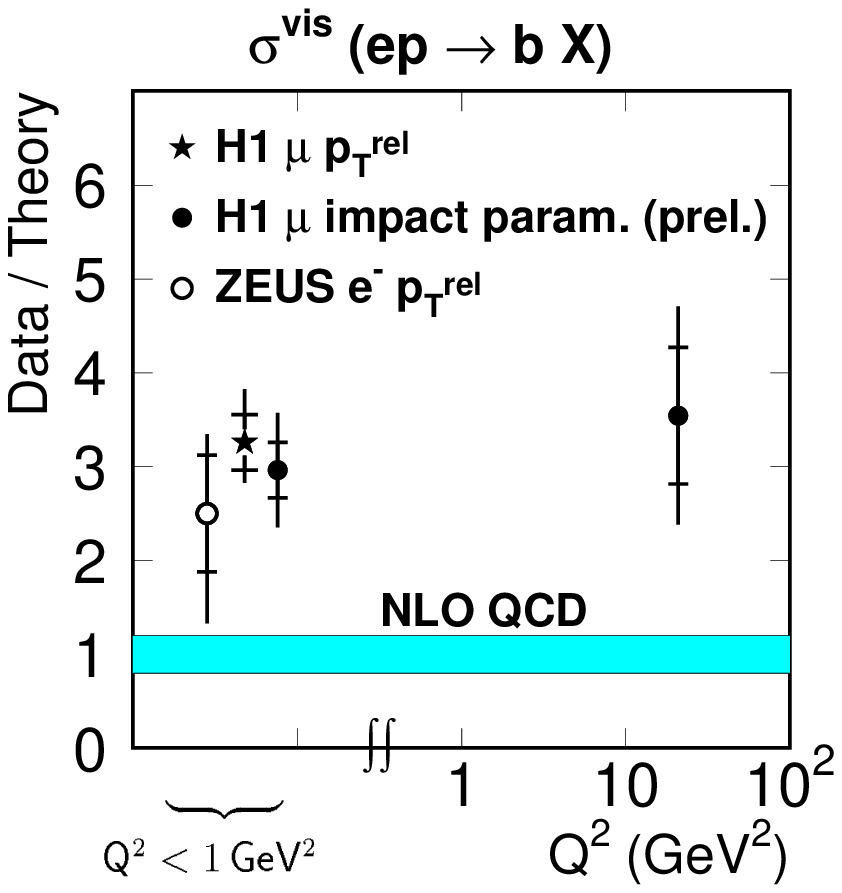}
\caption{ Left: ZEUS data for rapidity distribution for b in photoproduction 
in comparison to PYTHIA and CASCADE~\protect\cite{jung};\protect\\
Right: Comparison of measured b production cross section 
at HERA to NLO QCD calculation as function of 
$Q^2$~\protect\cite{turcato}. \label{jung-fig4}}
\end{figure}

In DIS$_{\mathrm {ep}}$ events the measurements of H1 and ZEUS for D$^*$
production for 
the total cross section and the differential distributions in
$p_T^{\mathrm{D*}}$ and $\eta^{\mathrm{D*}}$
are nicely described by CASCADE.
 The HVQDIS Monte-Carlo,
traditionally used in ep-scattering for heavy flavour production, 
fails in describing the forward
($\eta_{\mathrm{D*}}$) range giving 
too low cross section (Fig.~\ref{west-fig2}-right), similarly problem is seen 
 in photoproduction  events (Fig.10-left).

 The total cross section for b production  
has been studied both in photoproduction,
and in DIS$_{\mathrm {ep}}$ events.
ZEUS  results for  photoproduction
are presented in  Fig.~\ref{jung-fig4}-left for a distributions of the 
$\mu$ coming from b-quark decays. There is an 
agreement with CASCADE prediction, except in the forward direction. 
Here the $k_t$-factorization may break down~\cite{jung}. 
Note, that in this region
 $x_{\mathrm p}$ is here rather large, while $x_{\gamma}$ is small.

The visible cross section and 
the total cross section from  Monte Carlo simulations 
have been used to convert the b quark cross section from restricted 
$p_{\mathrm T}$ and $\eta$ ranges to form the $Q^2$ and $y$ distributions. 
The standard (DGLAP) NLO QCD calculations
give  results being lower 
than the measurements as shown in Fig.~\ref{jung-fig4}-right; 
CASCADE is above these calculations, 
but still somewhat below the data~\cite{turcato}.

\subsection{Inclusive J/$\psi$ Production}

\begin{figure}[t]
\hspace*{-0.25cm}
\epsfxsize=0.55\textwidth
\epsfysize=0.55\textwidth 
\epsfbox{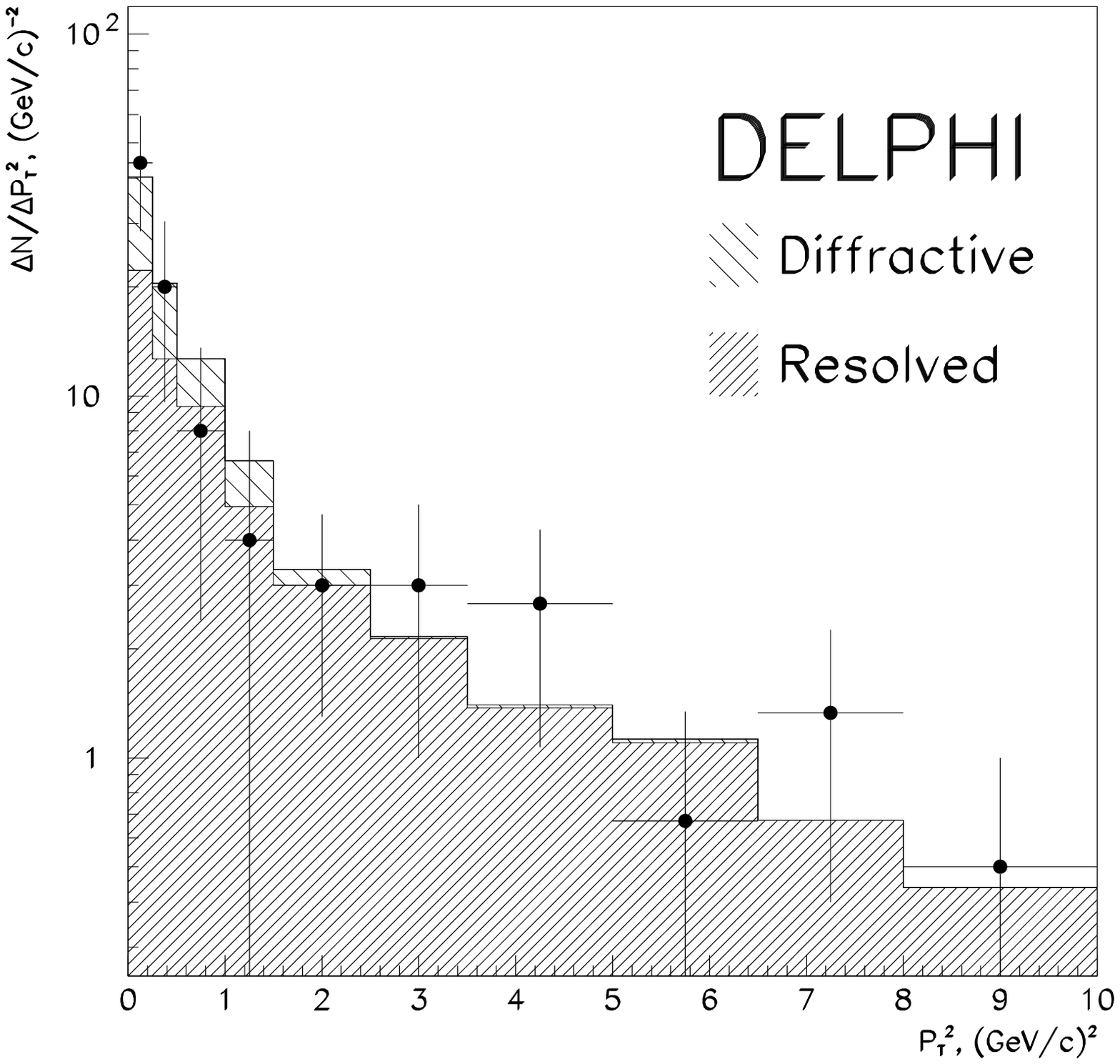}
\hspace*{-0.75cm}
\epsfxsize=0.55\textwidth
\epsfysize=0.55\textwidth
\epsfbox{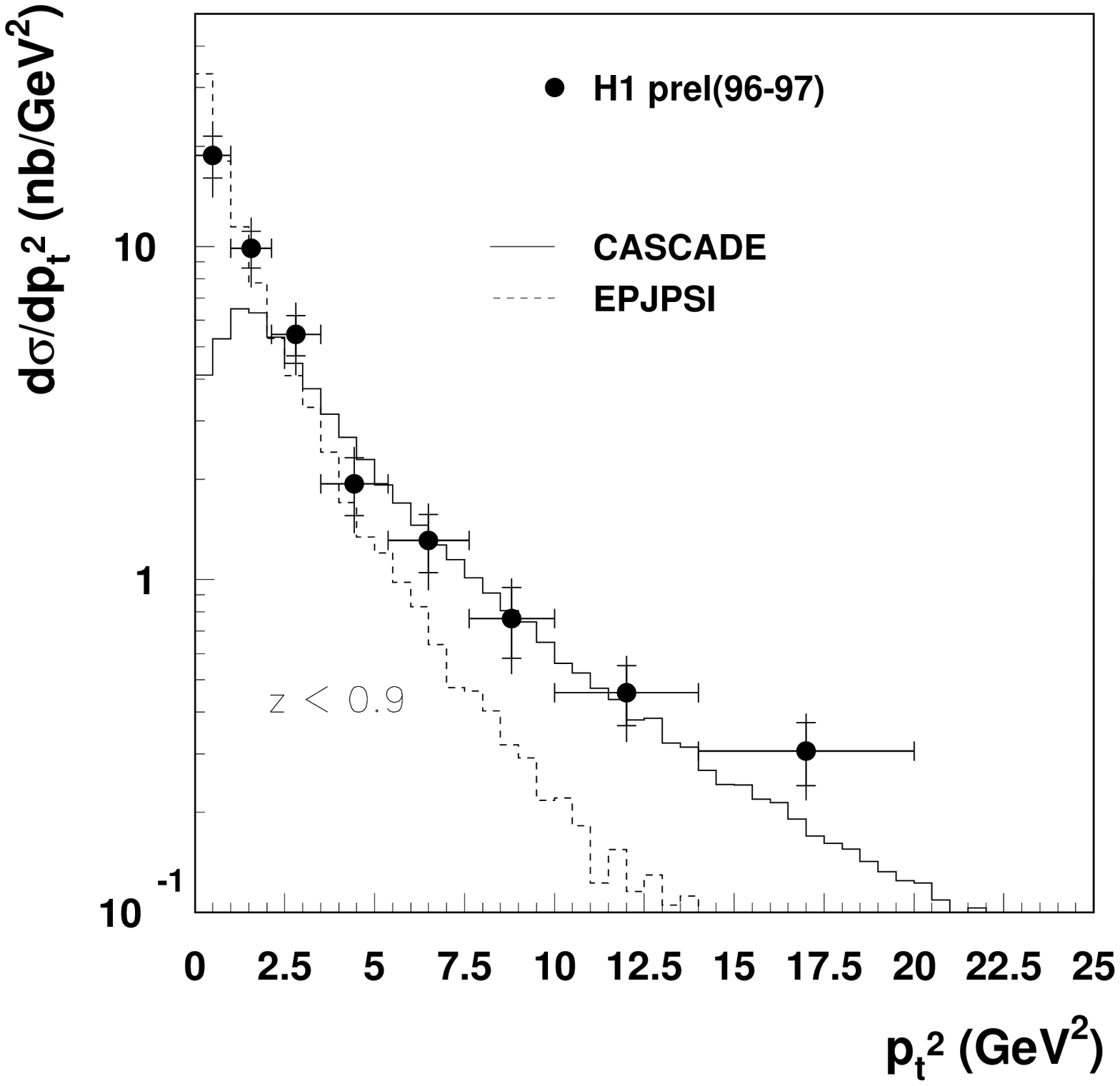}
\caption{Left: DELPHI data for differential cross section ${\mathrm d}\sigma/
{\mathrm d}p_T^2({\mathrm J}/\psi)$~\protect\cite{sokolov};\protect\\
Right: H1 data for ${\mathrm d}\sigma/{\mathrm d}p_T^2({\mathrm J}/\psi)$ in 
compared to MC~\protect\cite{jung}. \label{sokolov-fig12a}}
\end{figure}

In two-photon collisions at LEP a first look to $\mu^+\mu^- + {\mathrm X}$ 
has been given, where the muons are the decay products of the J/$\psi$,  
as shown in Fig.~\ref{sokolov-fig12a}~\cite{sokolov}-left. 
DELPHI finds $36 \pm 7$ events in the muonic decay. The $p_T$ spectrum 
allows for the separation of the resolved (gluon exchange) and 
diffractive process (pomeron exchange). The latter fraction is found 
to $26 \pm 22\%$. A precise measurement may help understand the 
total $\gamma\gamma$ cross section. Recently~\cite{klasen}, it was 
proposed that a comparison with theoretical calculations of 
J/$\psi$ plus two jets 
may show, if colour octet production is important as it was  found at 
Tevatron and at LEP in ${{\mathrm e}^+{\mathrm e}^-}$ annihilation events.

In continuum events taken by BaBar (SLAC) 
close to the $\Upsilon (4S)$ both cross section and 
polar angular distribution of inclusive J/$\psi$ production 
favour NRQCD fragmentation with the colour octet model~\cite{passaggio}.

 The production of J/$\psi$ at HERA (H1) is also shown in
 Fig.~\ref{sokolov-fig12a}-right in agreement with the CASCADE model.

\section{Total Cross Sections for $\gamma \gamma$ and $\gamma^* \gamma^*$.  
Diffraction}
Pure perturbative QCD can not  describe  soft and 
semi-hard inclusive processes at very high energy. The  old, 
very well known and successful framework for soft phenomena is  based on 
 the Regge   models, with exchange of Pomeron and Reggeons. 
The Pomeron is a complicated object with vacuum quantum numbers;
its exchange 
governs  the elastic and diffractive processes at high energies or at
 small $x$. Other processes and quantities, such as the difference of 
the total cross sections for pp and ${\mathrm {p\bar p}}$ collisions,  
is thought to be related to 
the exchange of the Odderon, a partner of the Pomeron but with opposite
 $C$ and $P$ quantum numbers.
 There are various attempts to apply   the perturbative  QCD to the small $x$
regions. Simple models based on the  two-gluon and three-gluon
exchange  to represent Pomeron  and Odderon exchange. 

Important progress has been 
obtained recently in this field, new theoretical and experimental 
results were presented~\cite{fadin,kim,pancheri,berndt}.
For a consistent description at small $x$ in the BFKL approach, 
the impact factors of $\gamma^*$ at NLO are needed~\cite{fadin}. 
The advantage of the impact factors is that they can be calculated from 
first principles in the perturbative QCD (the photon-Reggeon interaction).  

The only place at present to test Odderon exchange processes is the HERA 
collider. A dedicated study of photoproduction with exclusive production of mesons 
decaying into  photonic final states - with even or odd number of photons 
(from two to five photons) - 
has been performed by H1 and no sign of an Odderon contribution was 
found~\cite{berndt}.

\subsection{Cross Section $\sigma_{\gamma \gamma}$} 
The reliable description of the process 
$\gamma \gamma \rightarrow \rm hadrons$
is very important not only from the theoretical point of view, 
but also since this process gives  the bulk of background 
at future linear colliders. This  cross section, $\sigma_{\gamma \gamma}$,
 can be related to the total cross sections for hadron production 
in collisions: pp, ${\mathrm {p\bar p}}$, and $\gamma$p since all of them are 
 dominated  by the exchange of the Pomeron.
Assuming the factorization  and the Additive Quark Model the  
data for   all these processes, after the appropriate rescaling,
should lie on a universal curve. This is not observed
- see Fig.~\ref{pancheri-tot}-left, where a problem with the normalization of $\gamma \gamma$ data is indicated~\cite{pancheri}. 
The high energy $\gamma \gamma$  data were checked to be
 consistent with each other, see Fig.~\ref{pancheri-tot}-right,
 where all LEP data (together with PLUTO, TPC-2$\gamma$ and MD-1 results) 
 are presented as a function of $W_{\gamma \gamma}$ 
\cite{kienzle}.  
The  observed rising behaviour of cross sections
can be obtained in e.g., minijet models. To ensure unitarity 
the Eikonal Minijet Model 
 was applied to describe the data 
(Fig.~\ref{kienzle-sig}-left) and  it would appear that data for 
$\sigma(\gamma \gamma \rightarrow \rm hadrons)$ are overestimated by 
about 10\%. Effects of resummation were also studied~\cite{pancheri}.

A DELPHI measurement of the $\sigma_{\gamma^* \gamma^*}$
for very low $Q^2$ double-tag events,
important for a reliable estimation of the $\sigma_{\gamma \gamma}$,
  was also presented~\cite{nygren}, 
see Fig.~\ref{kienzle-sig}-right for results.

\begin{figure}[t]
\epsfxsize=0.55\textwidth
\epsfysize=0.55\textwidth 
\epsfbox{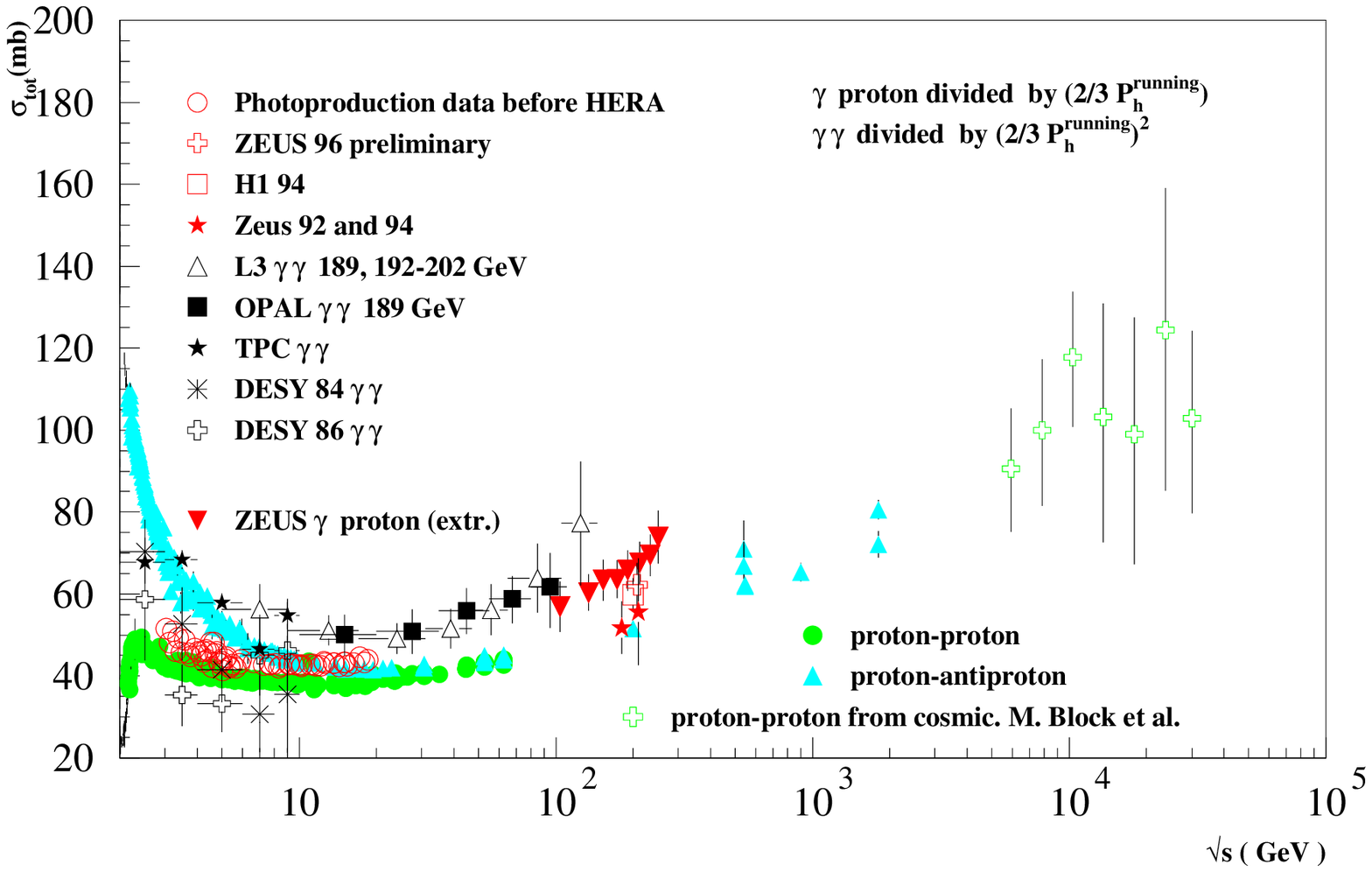}
\epsfxsize=0.45\textwidth
\epsfysize=0.45\textwidth 
\epsfbox{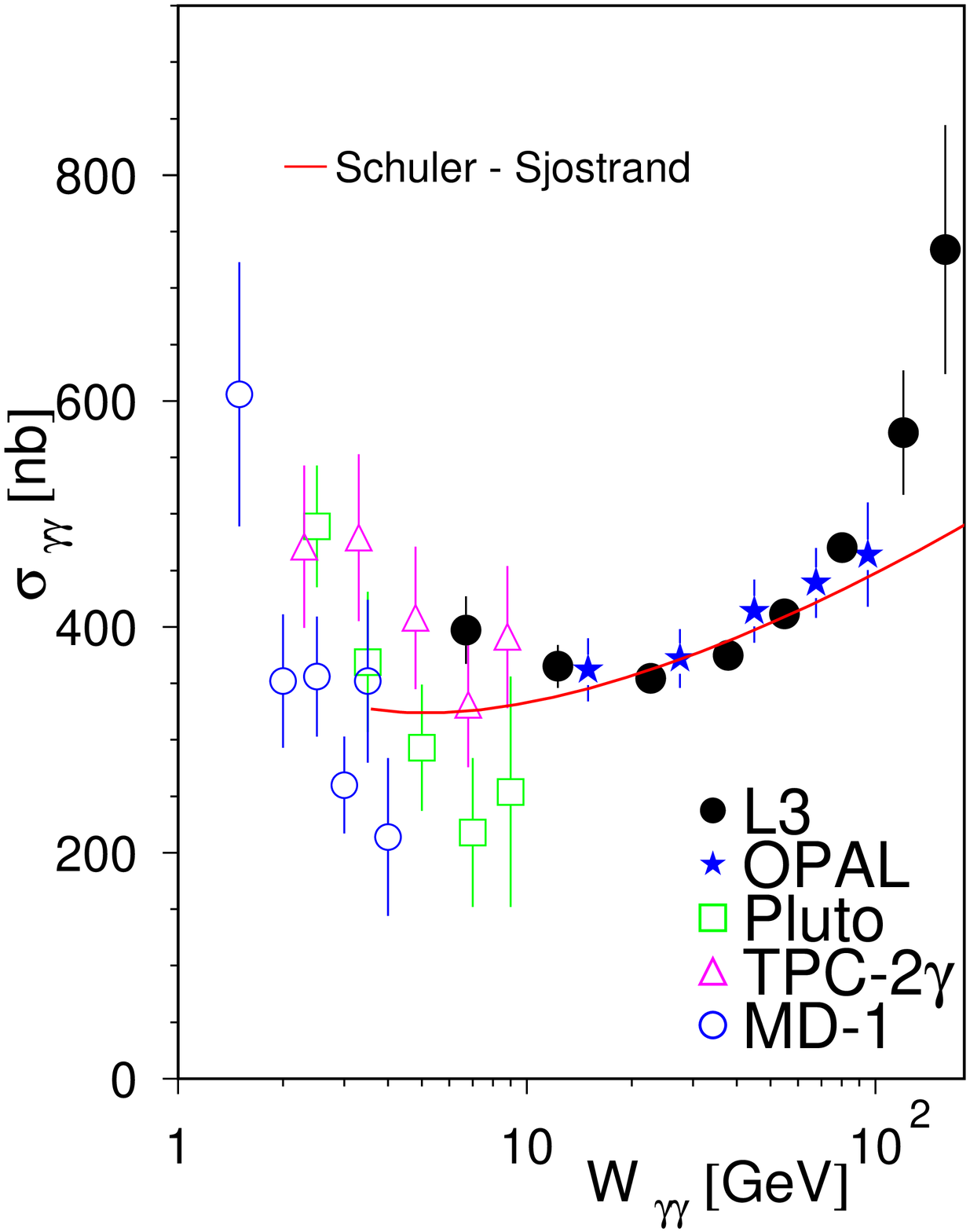}
\caption{Left: Measured total cross sections for pp, ${\mathrm {p\bar p}}$, 
and (rescaled) $\gamma$p, $\gamma \gamma$~\protect\cite{pancheri};\protect\\
Right: LEP data for $\sigma_{\gamma \gamma}$~\protect\cite{kienzle}. 
\label{pancheri-tot}}
\end{figure}
\begin{figure}[t]
\epsfxsize=0.55\textwidth
\epsfysize=0.5\textwidth 
\epsfbox{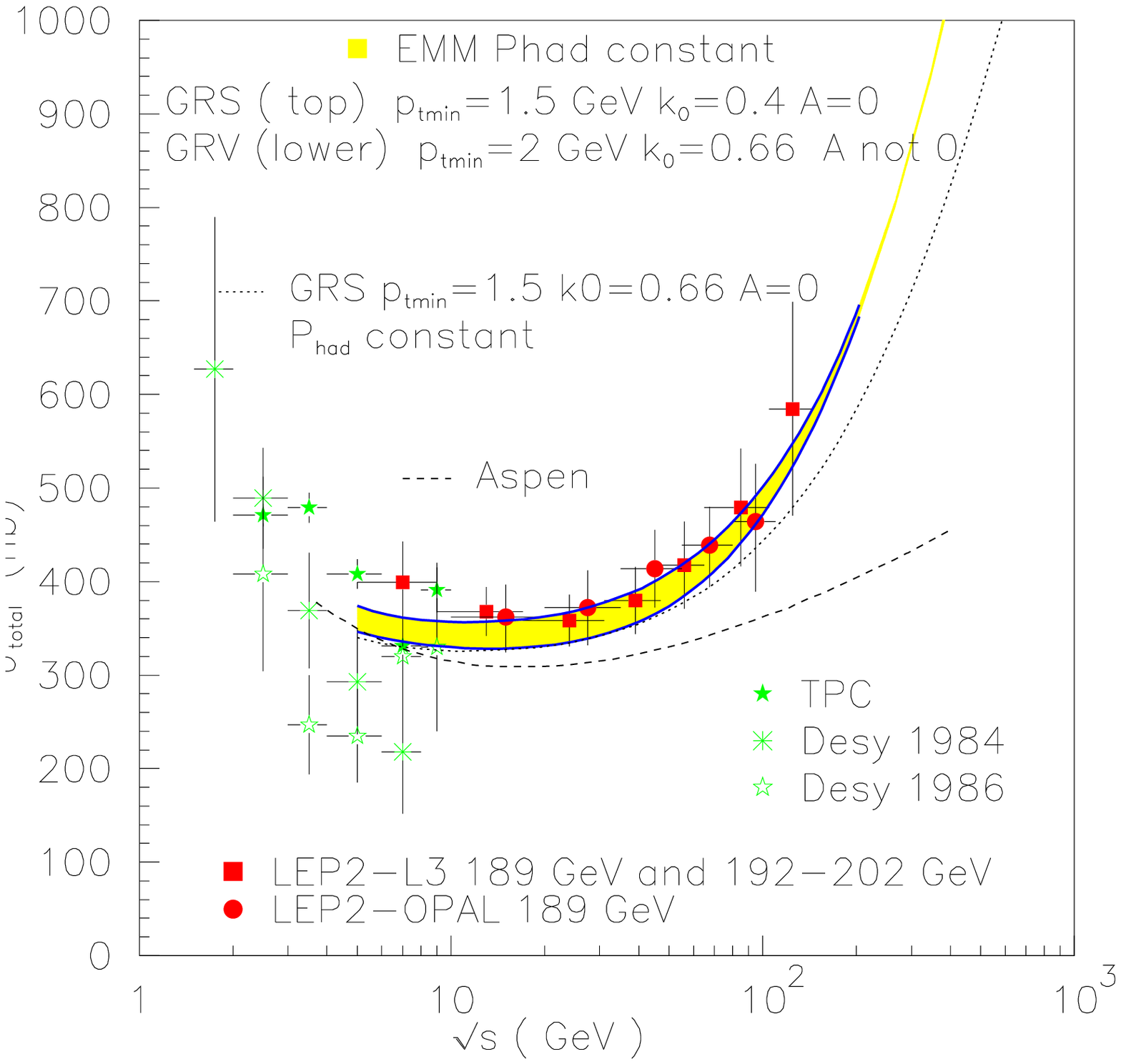}
\epsfxsize=0.49\textwidth
\epsfysize=0.49\textwidth
\epsfbox{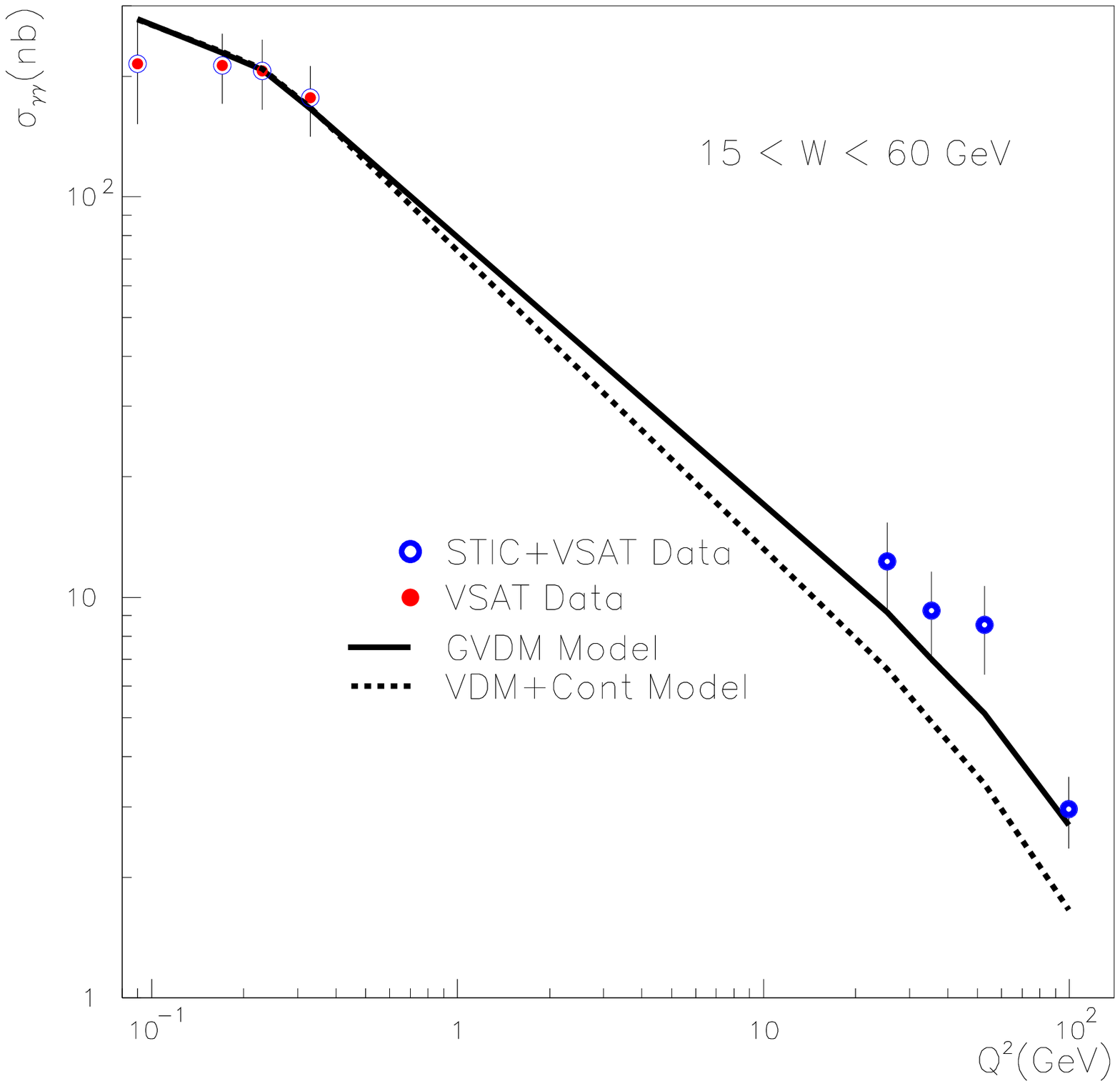}
\caption{Left: Two-photon cross section $\sigma_{\gamma \gamma}$ versus 
$\sqrt{s}$ in various models~\protect\cite{pancheri};\protect\\
Right: The DELPHI data for very low $Q^2$ 
double-tag events~\protect\cite{nygren} \label{kienzle-sig}.}
\end{figure}
\subsection{Cross Section $\sigma_{\gamma^* \gamma^*}$}
In the virtual photon collisions, $\gamma^* \gamma^* \rightarrow \rm hadrons$,
 with large virtualities of both photons, $Q^2_1$$\sim Q^2_2$,
  a large scale  
 $Q^2_1$$\sim Q^2_2 \sim Q^2$, allows for a clean test of perturbative 
 QCD in Regge limit.   Such measurements  with the 
 aim to test the BFKL picture, as DGLAP evolution is suppressed for 
 $Q^2_1$$\sim Q^2_2$,   were performed by the 
LEP experiments~\cite{prange,linprzy}, 
see Fig.~\ref{prange-kim-bfkl}-left for results~\cite{prange}.
It is shown that LO BFKL can not describe the data, however, when the NLO 
corrections to BFKL expressions in the BLM optimal 
scale setting were implemented~\cite{kim} a good agreement is found
(Fig.~\ref{prange-kim-bfkl}-right).
\begin{figure}[t]
\hspace*{-0.2cm}
\epsfxsize=0.485\textwidth
\epsfysize=0.485\textwidth 
\epsfbox{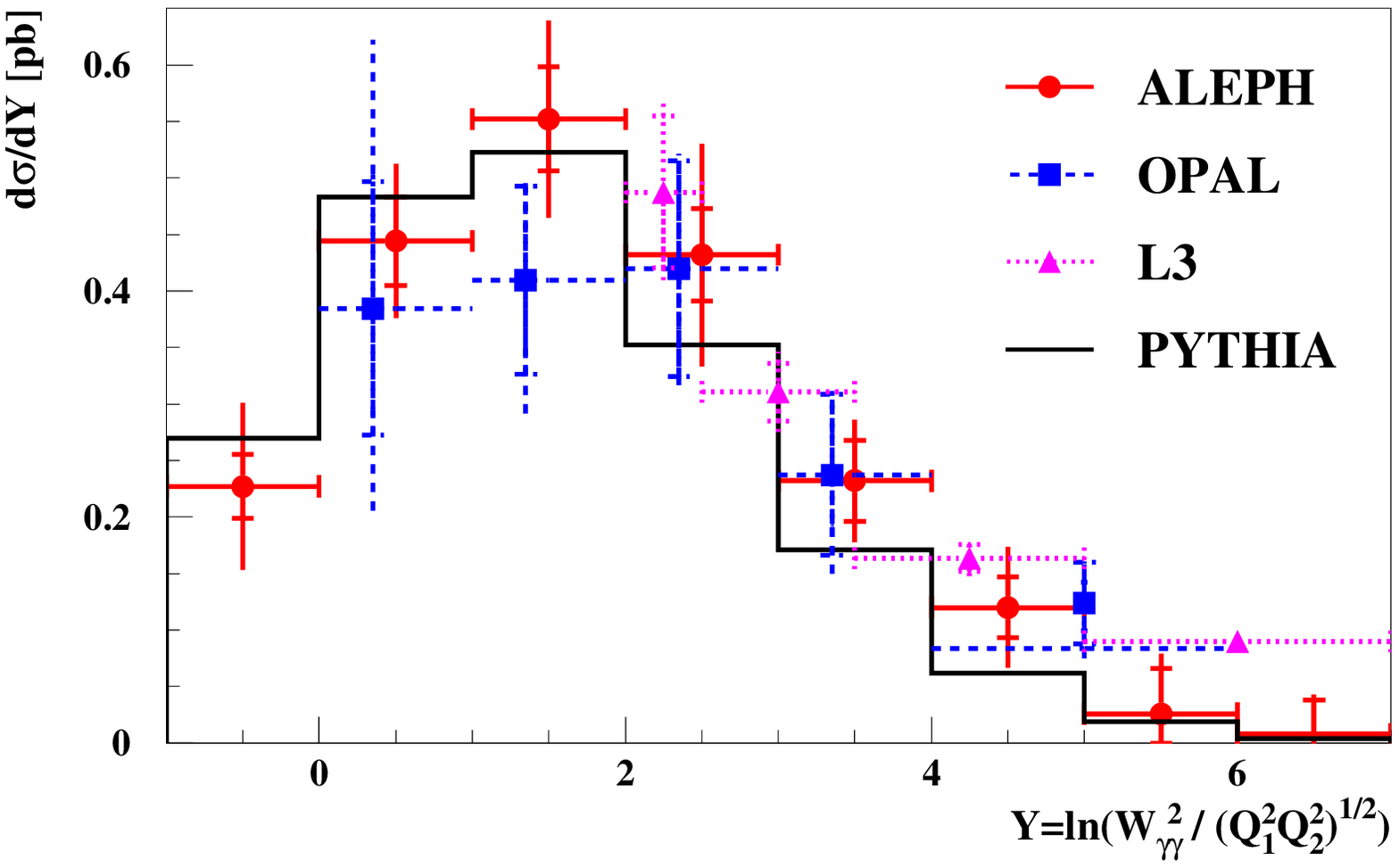}
\hspace*{0.2cm}
\epsfxsize=0.53\textwidth
\epsfysize=0.53\textwidth 
\epsfbox{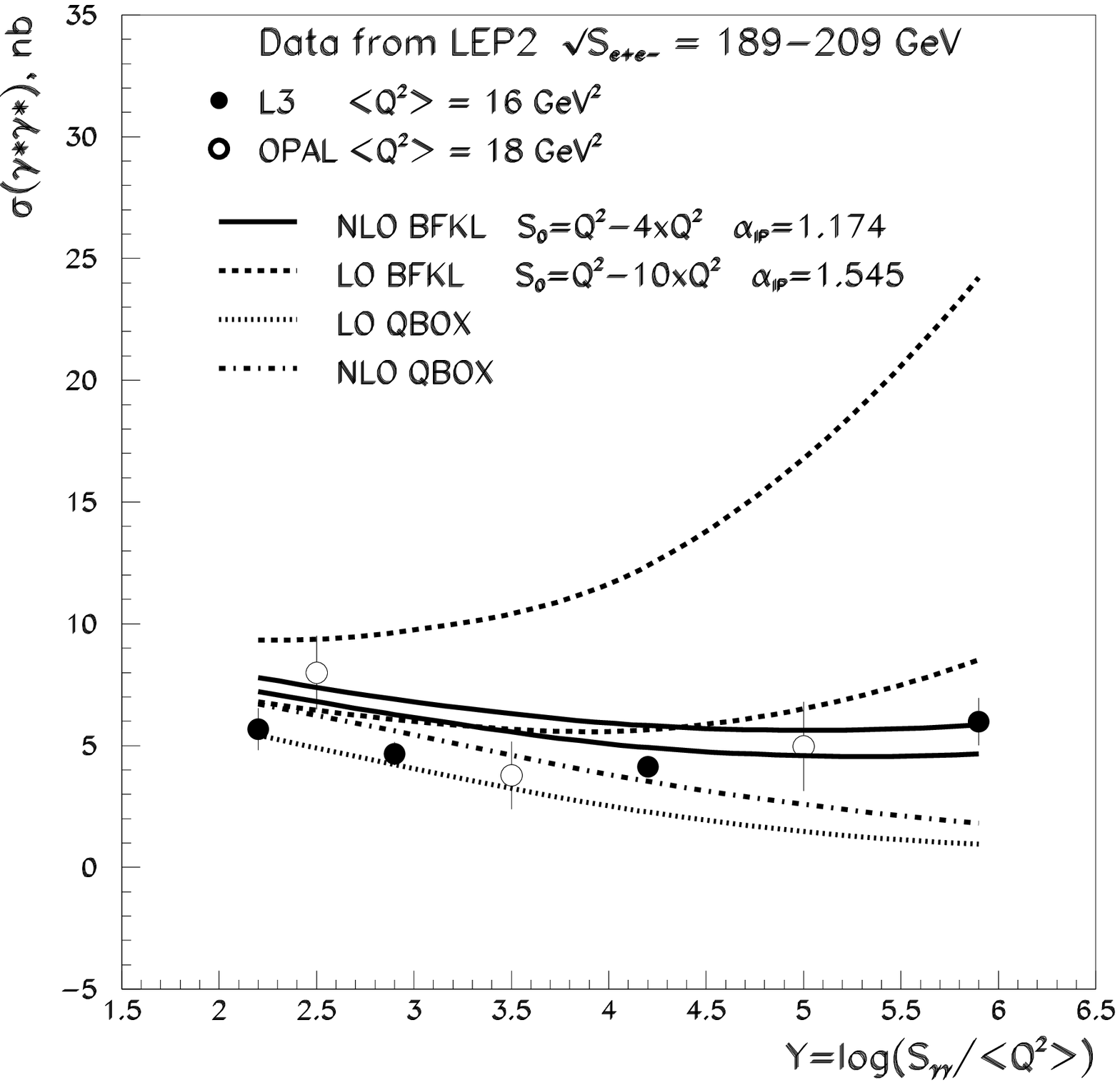}
\caption{Cross section for $\gamma^* \gamma^* \rightarrow {\mathrm {hadrons}}$ 
as function of $Y= \ln W_{\gamma \gamma}^2/Q^2$.\protect\\
Left: The LEP~2 data, by ALEPH, L3 and OPAL, in comparison to 
PYTHIA~\protect\cite{prange} (the OPAL and L3 data are extrapolated to 
the acceptance range of ALEPH),\protect\\ 
Right: Comparison of the LO and NLO BFKL predictions
to L3 and OPAL data~\protect\cite{kim} \label{prange-kim-bfkl}.}
\end{figure}
\subsection{Diffraction  at  Low and High Virtuality of the Photon at HERA.
 DVCS}
Various models of diffraction exist 
(e.g., pure soft Regge exchange models, the resolved Pomeron model~\cite{IG} 
 and dipole models~\cite{dipole}) and they can 
be tested by comparison of distributions in specific processes.
The diffractive production of vector mesons were measured in 
the photoproduction and DIS$_{\mathrm {ep}}$ events at HERA. 
The   photoproduction of vector mesons are shown in 
Fig.~\ref{kananov-fig21-rho}-left, 
 where  a steeper rise with energy is observed for heavy mesons as compared 
to light mesons~\cite{kananov}. 
For electroproduction the vector meson production ratios as a function 
of $Q^2$ were used to test the flavour independence hypothesis: 
$\rho:\omega:\phi:{\mathrm J}/\psi$=9:1:2:8, or 
(slightly modified ratios
 predicted in pQCD). 
For the light vector mesons the data scale with $(Q^2+m^2)$, 
while the ones for J/$\psi$ do not.
In Fig.~\ref{kananov-fig21-rho}-right the rate for the $\rho$ production 
as a function of $Q^2$ is presented in a form of 
 the ratio $R=\sigma_L$/$\sigma_T$, and found to be in agreement with 
QCD~\cite{kananov}.

The Deeply Virtual Compton Scattering (DVCS)
process, ${\mathrm {\gamma^* p \rightarrow \gamma p}}$, is an
 exclusive analogue to the prompt $\gamma$ production in DIS$_{\mathrm {ep}}$:  
${\mathrm {\gamma^* p \rightarrow \gamma X}}$. 
In DVCS scales are different: 
$Q^2$ is much larger than $(p_T^{\gamma})^2$, and $|t|$ is smaller 
than 1 GeV$^2$. 
This process can be described in terms of the skewed parton 
distributions, which can be treated as a 
generalization of the standard parton densities. The results of 
measurements at HERA (H1) for the DVCS  
cross section as a function of $Q^2$ and $W$ are 
presented in Fig.~\ref{dvcs}.
\begin{figure}
\epsfxsize=0.52\textwidth
\epsfysize=0.52\textwidth 
\epsfbox{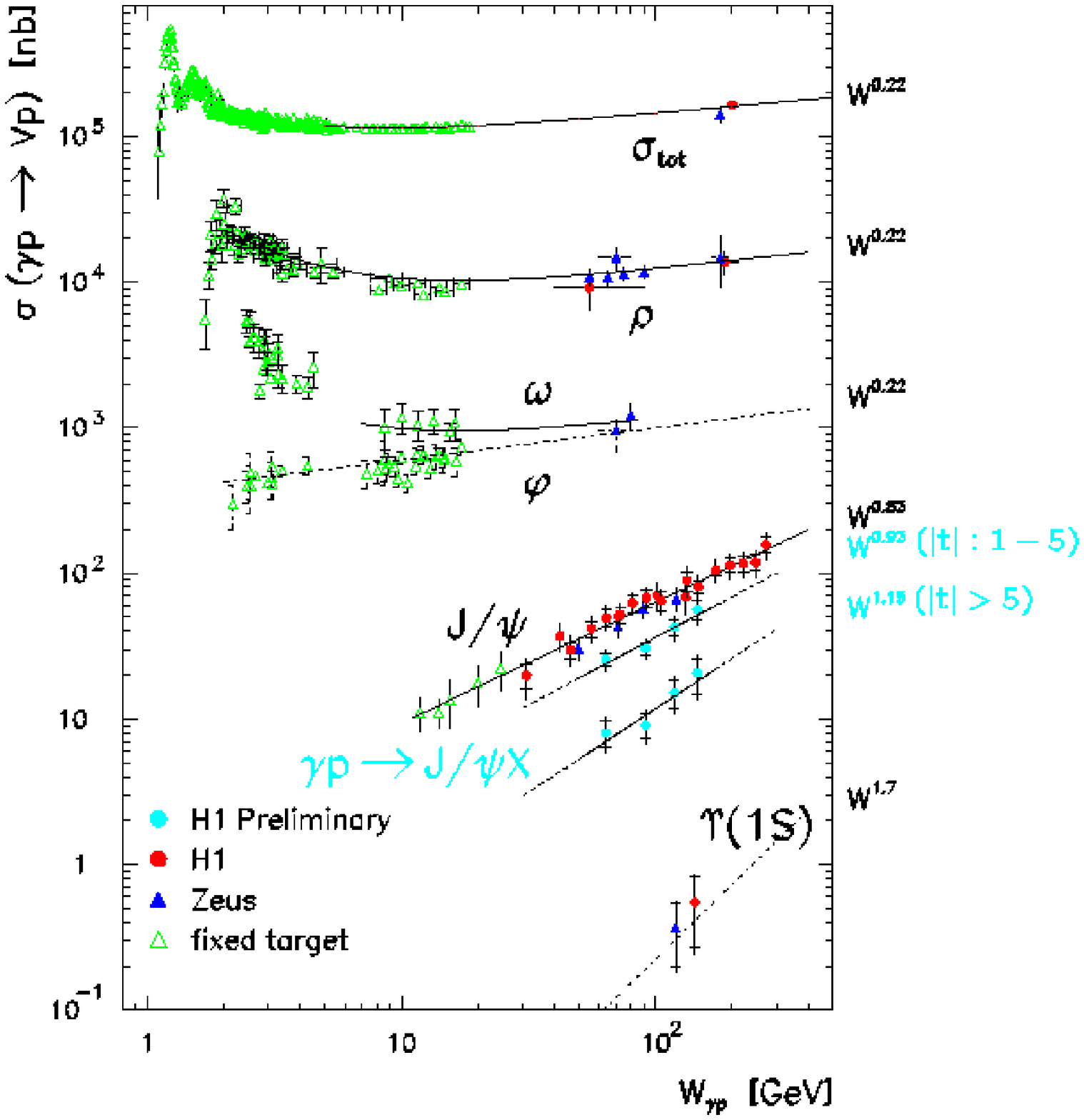}
\hspace*{-1.0cm}
\epsfxsize=0.55\textwidth
\epsfysize=0.55\textwidth 
\epsfbox{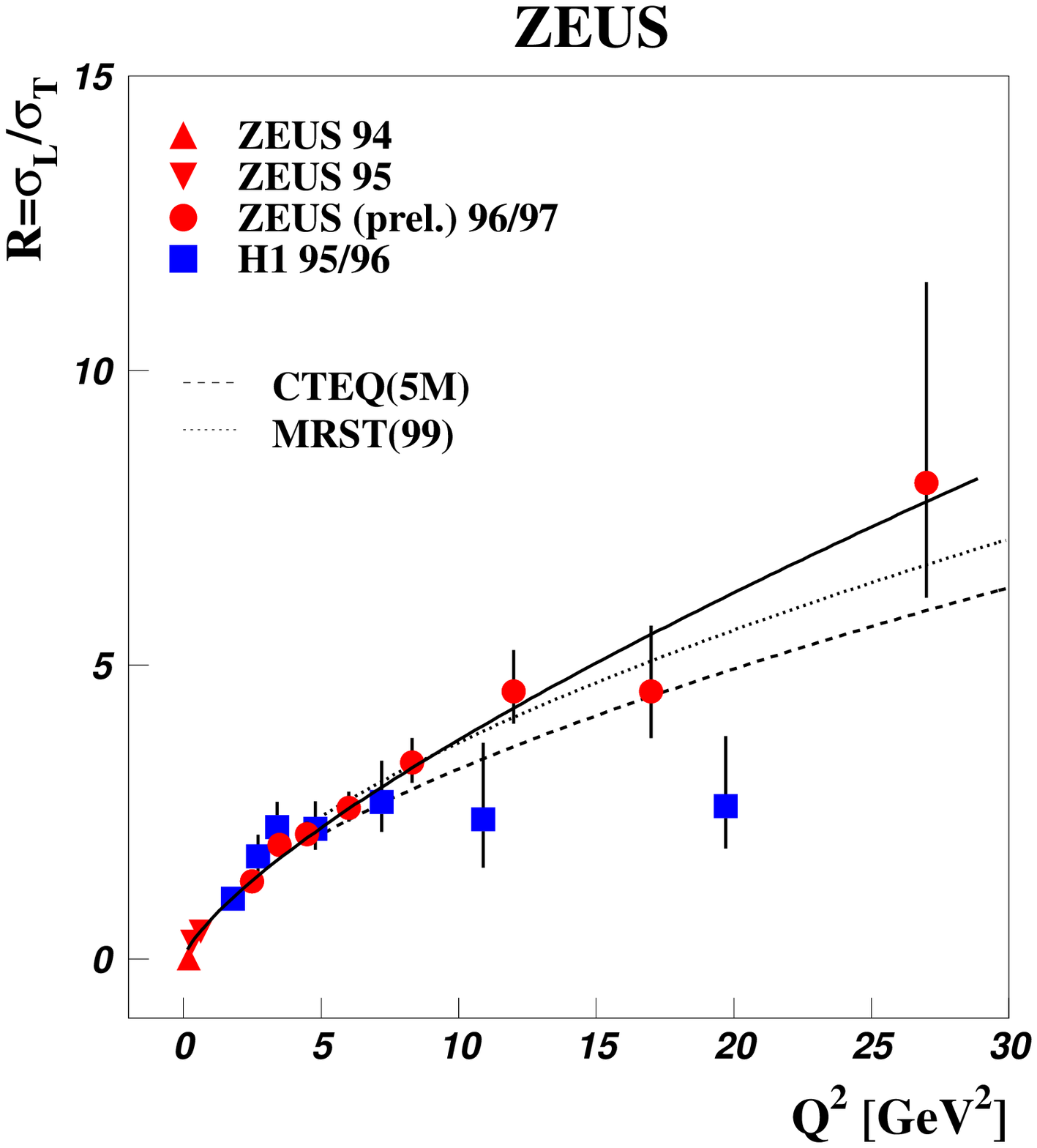}
\caption{Left: Photoproduction of vector mesons at HERA and 
fixed target exp.~\protect\cite{kananov},\protect\\
Right: The ratio $R=\sigma_L$/$\sigma_T$
for the $\rho$ production as a function of $Q^2$~\protect\cite{kananov}.
 \label{kananov-fig21-rho}}
\end{figure}

\begin{figure}
\begin{center}
\epsfig{file=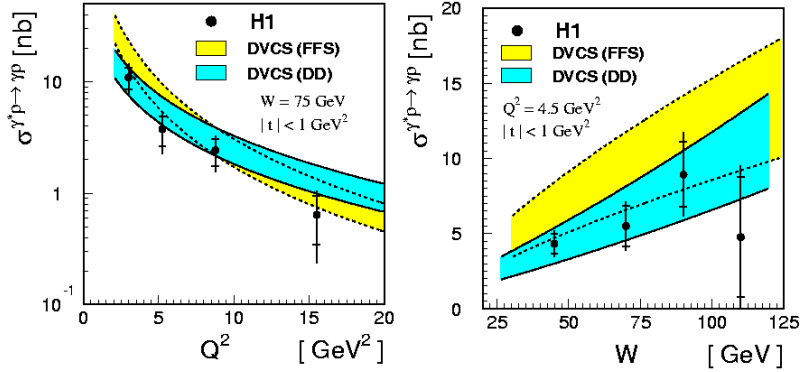,angle=90,width=0.9\textwidth}
\caption{Cross sections for DVCS as function of $Q^2$ and $W$, 
from H1~\protect\cite{kananov}.
 \label{dvcs}}
\end{center}
\end{figure}

In the HERMES experiment at HERA polarized e$^{\pm}$ beams collide 
with the (gas) targets. The exclusive diffractive 
$\rho$ production in  DIS$_{\mathrm {ep}}$ events was analyzed. 
The results for $\sigma(\gamma^*_L {\mathrm {p \rightarrow \rho p)}}$ are  
shown in Fig.~\ref{lipka-fig1} 
for two $Q^2$ samples ($\langle Q^2 \rangle \approx$ 2 and 
$4\unit{GeV^2}$). The quark exchange subprocess dominates
 for smaller $Q^2$ events, while gluon exchange mechanism starts to be 
important at larger $Q^2$. 
Spin effects in $\rho$ electroproduction were studied with the conclusion 
that s-channel helicity conservation is slightly violated. The  double-spin 
asymmetry were also studied~\cite{lipka} (for DIS and photoproduction)
 and results were interpreted  in terms of GVMD. 
\begin{figure}
\epsfxsize=0.6\textwidth
\epsfysize=0.45\textwidth 
\epsfbox{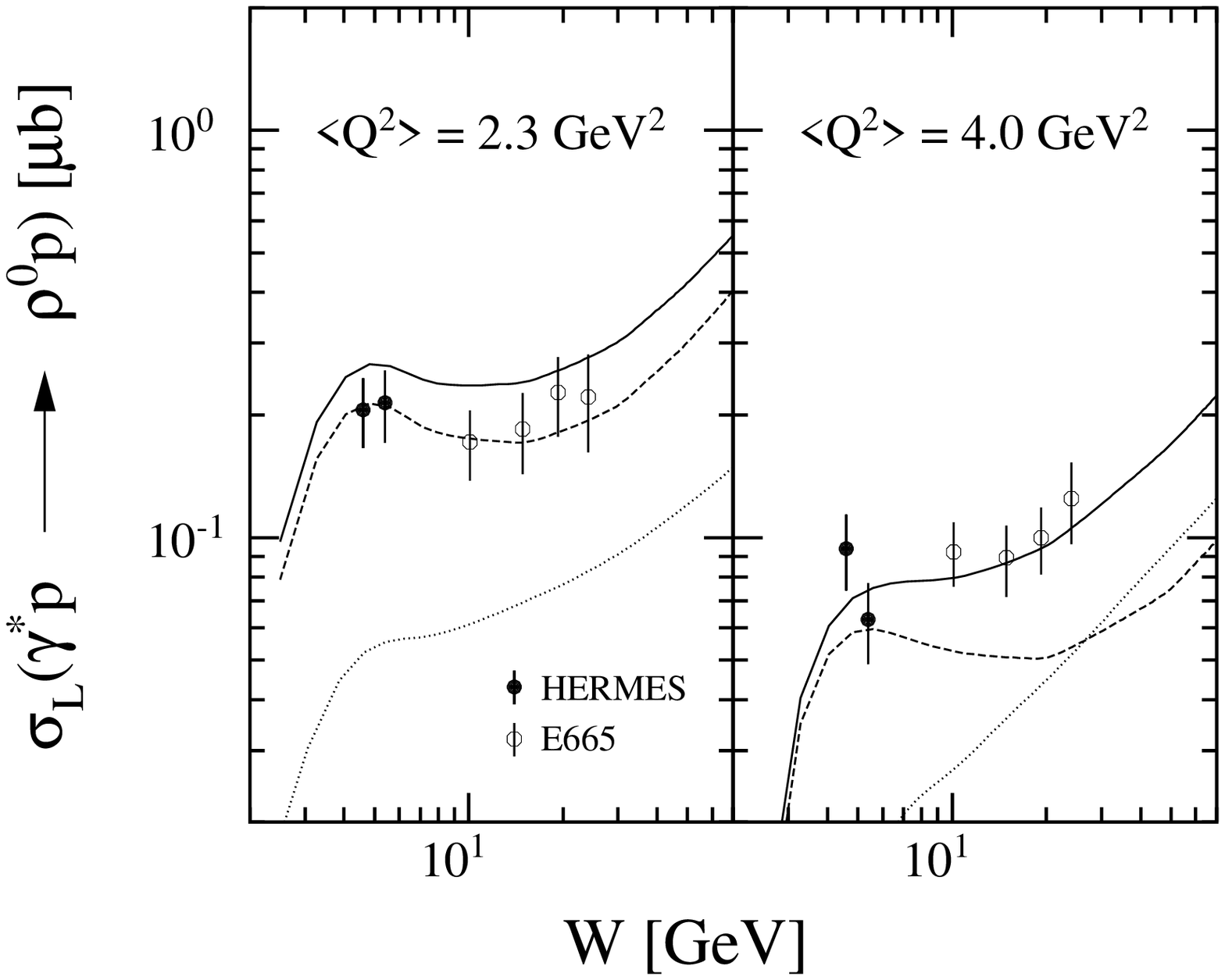}
\caption{HERMES data for $\rho$ production: 
$\sigma(\gamma^*_L {\mathrm {p \rightarrow \rho p}})$ 
for two $Q^2$ samples- dotted (dashed) line corresponds to the gluon (quark)
exchange subprocesses~\protect\cite{lipka}. \label{lipka-fig1}}
\end{figure}

\section{Exclusive Channels and Resonances in $\gamma \gamma$ collisions}

Study of pure leptonic final states in $\gamma \gamma$ collision 
were performed  to test 
QED (up to  ${\cal{O}}(\alpha ^4)$)
  but also to demonstrate the understanding of 
the detectors. 
The remarkable number of new results on   
exclusive hadronic final states presented during the 
conference, some of them from new high luminosity
low energy colliders, leads to better understanding of strong 
interaction at low energies. 
If a large momentum scale is involved in the reaction,  then we can use 
pQCD assuming factorization of   the  amplitude
 into a perturbative and non-perturbative part~\cite{brodsky-lepage}.
 In Refs.~\cite{kroll} the meson distribution amplitude were constrained 
 from study of transition $\gamma^* \gamma \rightarrow \pi,\eta,\eta'$. 
Radiative decay amplitudes for various particles, among them the decays 
 $\pi,\eta,\eta' \rightarrow \gamma^* \gamma$, lead to 
reconstruction of pion and  photon wave functions~\cite{anisovich}. 

Production of particles containing heavy quarks is of special interest; 
 results obtained at ${\mathrm {e^+e^-}}$ colliders (see below) 
and at HERA for the production of excited P-wave charm 
mesons~\cite{karshon} were discussed during the conference.

One of the basic open question of QCD is the existence of the glueballs, a 
bound state of gluons. The two-photon process is a powerful anti-filter 
for glueballs. The lattice simulation~\cite{bali} of  the quenched glueball
spectrum  was preformed with first promising results for $n_f$=2. 
\subsection{Lepton and Hadron Pairs}
Lepton pair production ${{\mathrm e}^+{\mathrm e}^-} \rightarrow 
{{\mathrm e}^+{\mathrm e}^-} \ell^+ \ell^-$ ($\ell = \mu, \, \tau$) 
have been measured by L3 
at LEP~2 as function of centre-of-mass energy and 
of $W_{\gamma\gamma}$. Limits on a possible anomalous coupling of the 
$\tau$ have been given~\cite{debreczenihaas}.

\begin{figure}[t]
\epsfxsize=0.4\textwidth
\epsfysize=0.4\textwidth 
\epsfbox{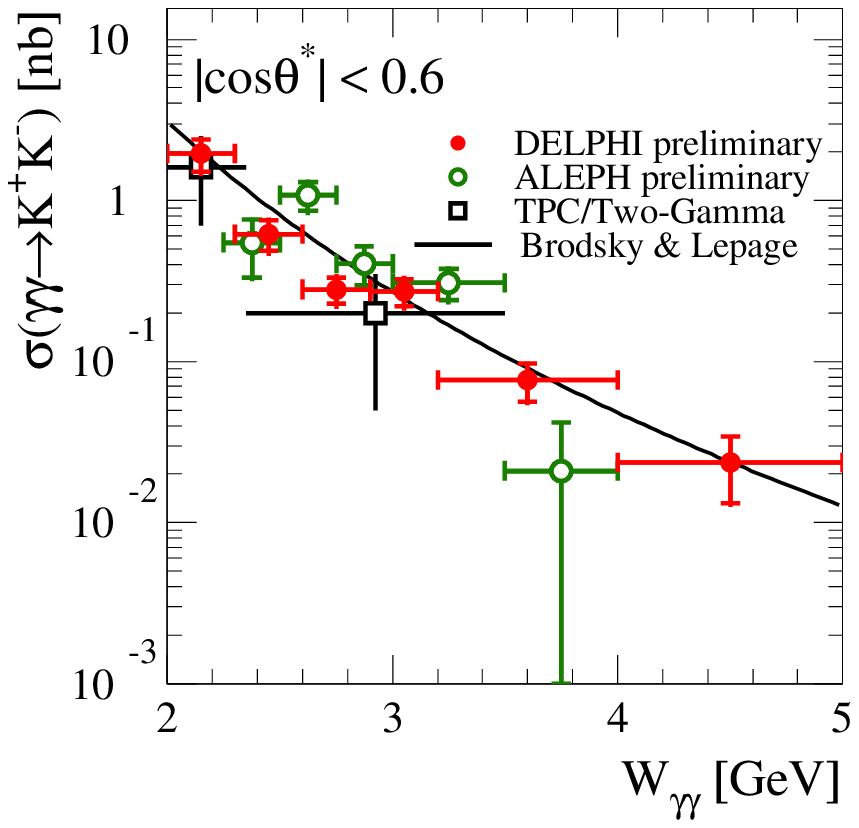}
\epsfxsize=0.4\textwidth
\epsfysize=0.4\textwidth 
\epsfbox{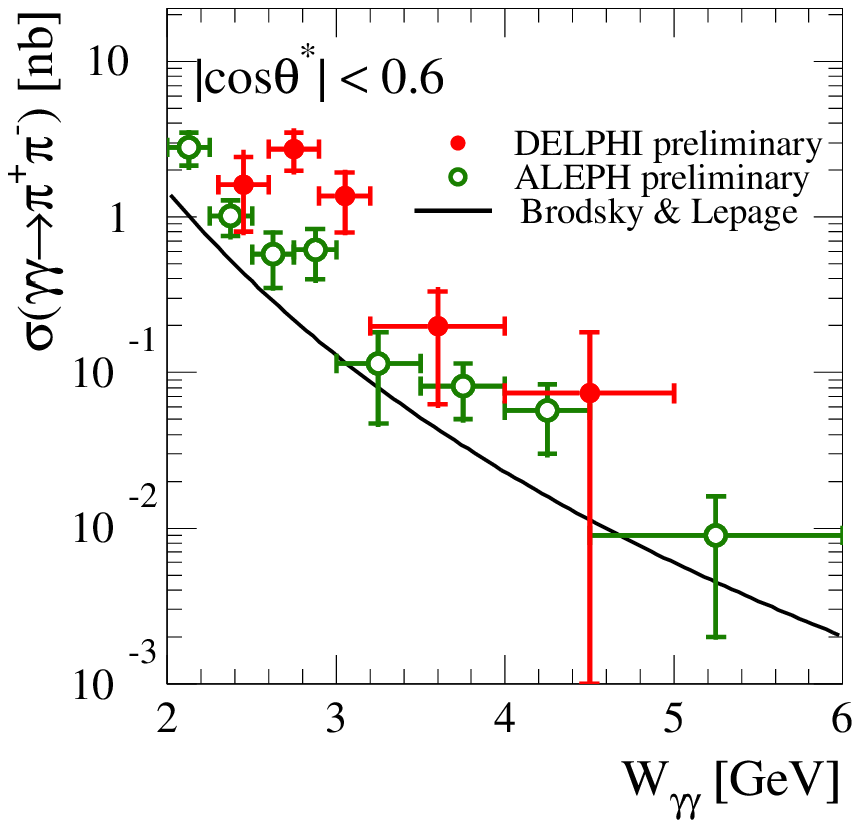}
\caption{Differential cross section for $\gamma\gamma$ to two charged 
kaon ond pion pairs, respectively as function of $W_{\gamma\gamma}$ 
as measured by DELPHI and ALEPH~\protect\cite{grzelak}. \label{grzelak-fig1}}
\end{figure}

New LEP~2 results on pion and kaon pair  
production were presented~\cite{grzelak} and   compared with 
the QCD-inspired calculation 
of Brodski and Lepage   
(Fig.~\ref{grzelak-fig1}). 
ALEPH and DELPHI find good agreement for kaon pairs both 
in normalization and shape of the distribution in scattering angle 
$\cos \theta ^*$ and $W_{\gamma\gamma}$. For the pions the shape of 
the distributions all agree, while for the normalization,
sensitive to nonperturbative input, neither 
the experiments nor experiments versus theory agree. 

The production mechanism for dibaryon production was tested by OPAL and Belle 
in ${{\mathrm p}\bar{\mathrm p}}$ and by L3 in $\Lambda\bar{\Lambda}$ final 
states. The dibaryon mass spectra as function of $W_{\gamma\gamma}$ 
(from the dimensional counting rule $\sim s^{-n}$ is expected) prefer 
the quark-diquark model ($n=4$) rather than the three-quark model ($n=6$). 
The measurements are in agreement with previous measurements, though all 
predictions tend to lie below the data~\cite{dibaryon}.
 
\subsection{Exclusive Meson Production and Glueball Search}
 The results are 
manifold and involved. The branching ratios, production as function 
of the virtuality of the photon and form factors of specific mesons 
are interesting on their own, but also constrain the gluonium content 
of the particle. The determination of the spin and helicity of 
resonances has been performed, showing that for the tensor mesons the 
multiplet is understood, while for the scalars the situation is more 
unclear. Possible mixture of states complicates the interpretation and 
the isolation of the best glueball candidate~\cite{amsler,braccini}.

\subsection{Charmonia and Bottomonia}

Extraction of $\Gamma_{\gamma\gamma}(\eta_{\mathrm c})$ was 
investigated in the $\pi^+\pi^-$K$^+$K$^-$, 2(K$^+$K$^-$),
K$_{\mathrm S}$K$^+ \pi^-$ and 2($\pi^+\pi^-$) decay modes by DELPHI. 
While the combination of the first three channel results in a 
two-photon width of $13.0 \pm 2.7 \pm 5.0 \unit{keV}$ and is 
compatible, though somewhat higher, with previous 
measurements, the non-observation of the 4$\pi$ decay leads to an 
upper limit of $3.8 \unit{keV}$~\cite{braccini}. 

The CLEO Collaboration presents a first observation of the
$\chi_{\mathrm c0}$ in two-photon collisions~\cite{paar}. A result of
$\Gamma_{\gamma\gamma}(\chi_{\mathrm c0}) = 3.76 \pm 0.65 \pm 1.81 \unit{keV}$
is extracted in the $\pi^+\pi^-\pi^+\pi^-$ decay mode. Belle shows that
the $\chi_{\mathrm c0}$ is present in the K$_{\mathrm S}$K$_{\mathrm S}$
mass spectrum as well~\cite{uehara}. 

\begin{figure}[t]
\begin{minipage}{.40\textwidth}
\epsfxsize=1.00\textwidth
\epsfysize=1.00\textwidth
\epsfbox{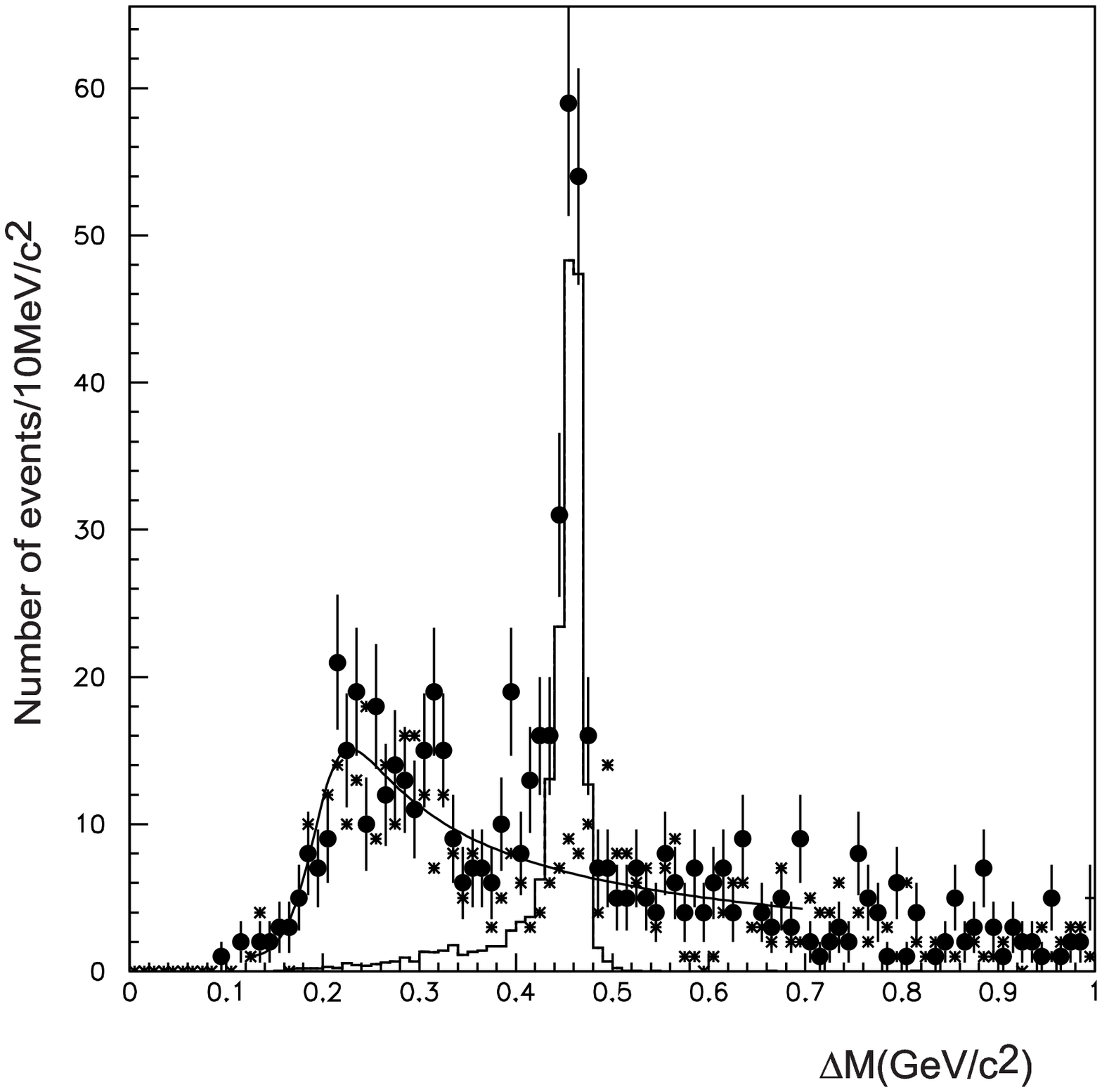}
\caption{Mass difference $m(\ell^+\ell^-\gamma) - m(\ell^+\ell^-)$ 
for $\chi_{\mathrm c0} \rightarrow {\mathrm J/\psi}$ events as measured by 
Belle~\protect\cite{uehara}. \label{uehara-fig3a}}
\end{minipage}
~
\begin{minipage}{.57\textwidth}
\epsfxsize=1.00\textwidth
\epsfysize=0.66\textwidth 
\epsfbox{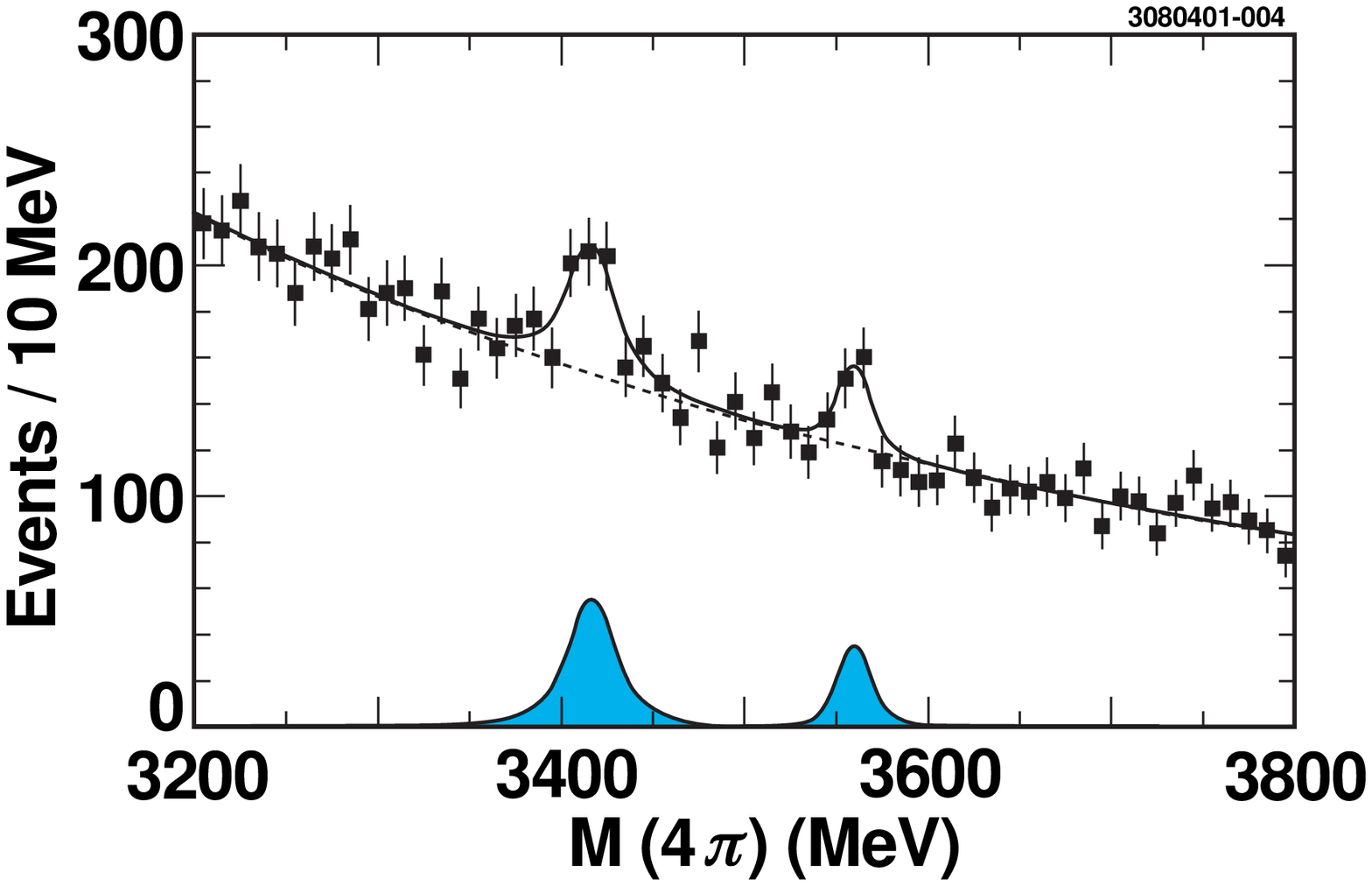}
\caption{Invariant mass distribution of four charged pions as measured 
by CLEO~\protect\cite{paar}. \label{paar-fig1}}
\end{minipage}
\end{figure}

Belle also measures the $\chi_{\mathrm c2}$ production in the 
J/$\psi \gamma$ decay mode (Fig.~\ref{uehara-fig3a}), CLEO presents a 
first measurement from the $\pi^+\pi^-\pi^+\pi^-$ decay mode 
(Fig.~\ref{paar-fig1}). Both results are lower, but in agreement 
with previous results in two-photon collisions and larger than the results 
from ${{\mathrm p}\bar{\mathrm p}}$ collisions. As pointed out~\cite{paar} the 
ratio $\Gamma_{\gamma\gamma}(\chi_{\mathrm c0}) / 
\Gamma_{\gamma\gamma}(\chi_{\mathrm c2})$ is a good QCD test 
with small theoretical uncertainties and could provide 
a measurement of $\alpha_{\mathrm S}$, but the calculations of the 
${\cal{O}}(\alpha_{\mathrm S}^2)$ term for the $\chi_{\mathrm c}$ 
states are badly needed. 

A first search for the $\eta_{\mathrm b}$ meson in 
two-photon collisions was performed in 
ALEPH~\cite{boehrer}, the bottomonium ground state being still 
undetected. The observation of one $\eta_{\mathrm b}$ candidate 
($m = 9.30 \pm 0.02 \pm 0.02 \unit{GeV}/c^2$) compatible with 
background gives upper limits of $57\unit{eV}$ and $128\unit{eV}$ for 
$\Gamma_{\gamma\gamma}(\eta_{\mathrm b}) \times$BR for decays to 
4 and 6 charged particles, respectively.

\section{Future Projects and Related Topics}
The two-photon collisions  at the future LHC collider
 may turn out to be important 
for the search of the Higgs boson, as addressed 
in the presentation~\cite{piotrzkowski}. 
The coherent photon-pomeron and photon-photon interactions appear already
in ultraperipheral collisions at RHIC;
a clear $\rho^0$ signal was observed in the $\pi^+\pi^-$ invariant mass 
spectrum~\cite{meissner}.

Relativistic nuclear collisions at LHC and RHIC were studied
in~\cite{serbo}. In particular  the Coulomb and unitarity
corrections to the single ${\mathrm {e^+e^-}}$ pair production as well as
the cross section $\sigma_n$ for the multiple pair production 
 were obtained in an analytic form. Some of the results differ   
from results published by other authors. 
 The Born cross section for ${\mathrm {e^+e^-}}$ pair production is given by:
$$\label{Landau}
\sigma_{\rm {Born}}= 
                 \frac{28}{27\pi}\,\frac{(Z_1Z_2\alpha^2)^2}{m_{\mathrm e^2}}
\left[L^3-2.198\,L^2+3.821\,L-1.632\right] \,, $$
with  $L=\ln(\gamma_1\gamma_2)\,$, where $\gamma_i$ and $Z_i$ are the Lorentz 
factors and the charges of the colliding nuclei. 
The  cross sections for this process at LHC and RHIC  are huge, 
so the pair production can be a serious background.
The control of this background may turn out important for a good 
beam lifetime and luminosity of the colliders. In Table 1 
results for $\sigma_{\mathrm {Born}}$ are given together with  corrections 
(calculated for small $1/L$ in the main logarithmic approximation).\\  
{\renewcommand{\arraystretch}{1.5}
\begin{table}[h]
\caption{Results for relativistic heavy-ion colliders ($Z_1=Z_2\equiv Z$ and 
$\gamma_1=\gamma_2\equiv \gamma$)}
\begin{center}
 \begin{tabular}{|c|c|c|c|c|c|}\hline
Collider & $Z$ & $\gamma$ & $\sigma_{\rm Born}$ [kb] &
$\frac{\displaystyle\sigma_{\rm Coul}}{\displaystyle \sigma_{\rm
Born}}$ & $\frac{\displaystyle\sigma_{\rm unit}}{\displaystyle
\sigma_{\rm Born}}$ 
\\ \hline
RHIC, Au-Au & 79 & 108 & 36.0 & $-25$\% & $-4.1$\%
\\ \hline
LHC, Pb-Pb & 82 & 3000 & 227 & $-14$\% & $-3.3$\%
\\ \hline
LHC, Ca-Ca & 20 & 3700 & 0.872 & $-1.06$\% & $-0.025$\%
 \\ \hline
\end{tabular}
\end{center}
\end{table}}
 
Very important for the investigations and the understanding of the photon 
and its interactions are  planned ${\mathrm {e^+e^-}}$ 
Linear Colliders (LC)
at large CM energy (500, 800 GeV or higher), in particular the 
Photon Collider (PC) option based on 
 Compton backscattering of laser light off the high energy electrons. 
Such  $\gamma \gamma$ and $e \gamma$ collider has unique properties
as it was discussed in Refs.~\cite{telnov,ginzburg} 
(with particular emphasis on the TESLA 
PC~\cite{TESLATDR,PC2000}). 
Cross sections are higher than for corresponding processes in 
${\mathrm {e^+e^-}}$ 
collision, accessible masses are larger and  there are unique reactions, 
such as $\GG\ \to$H. 
Since the \GG\ luminosity in the high
energy part of spectra at TESLA can be about 30 \% of the \EPEM\
luminosity (see also Ref.~\cite{arteaga} for a fast luminosity measurements
study)
  the number of
``interesting'' events at the photon collider will be even higher than
in \EPEM\ one.
A short list of physics processes
for the photon collider is presented in Table~\ref{processes}~\cite{PC2000}.

\begin{table}[hbtp]
\caption{Gold-plated processes at photon colliders}
\vspace{0mm}

{\renewcommand{\arraystretch}{1.0} \small
\begin{center}
\begin{tabular}{ l  c } 
\hline
$\quad$ {\bf Reaction} & {\bf Remarks} \\
\hline\hline
$\GG\to h_0 \to \bbbar, \GG$ & $M_{h_0}<160$ \GEV  \\
$\GG\to h_0 \to WW(WW^*)$    & $140<M_{h_0}<190\,\GEV$ \\
$\GG\to h_0 \to \ZZ(\ZZ^*)$      & $180<M_{h_0}<350\,\GEV$ \\
\hline
$\GG\to H,A \to \bbbar$  &
 \MSSM\ heavy Higgs \\

$\GG\to \tilde{f}\bar{\tilde{f}},\
\tilde{\chi}^+_i\tilde{\chi}^-_i,\ H^+H^-$ & supersymmetric particles \\ 
$\GG\to S[\tilde{t}\bar{\tilde{t}}]$ & 
$\tilde{t}\bar{\tilde{t}}$ stoponium  \\
$\GE \to \tilde{e}^- \tilde{\chi}_1^0$ & $M_{{\tilde{e}^-}} < 
0.9\times 2E_0 - M_{{\tilde{\chi}_1^0}}$\\
\hline
$\GG\to W^+W^-$ & anom. $W$ inter., extra dim. \\
$\GE^-\to W^-\nu_{e}$ & anom. $W$ couplings \\
$\GG\to WW+WW(ZZ)$ & strong $WW$ scattering\\
\hline
$\GG\to t\bar{t}$ & anom. $t$-quark interactions \\
$\GE^-\to \bar t b \nu_e$ & anom. $W tb$ coupling \\
\hline
$\GG\to$ hadrons & total \GG\ cross section \\
$\GE^-\to e^- X$ and $\nu_{e}X$ &  struct. functions \\ 
$\gamma g\to \qqbar,\ \ccbar$ & gluon distr. in the photon \\
$\GG\to J/\psi\, J/\psi $ & QCD Pomeron \\
\hline
\end{tabular}
\end{center}
}
\label{processes}
\end{table}
The unique feature of a \GE\ collider is the possibility  of
the measurement of the structure function
for the real photon (not quasi-real as presently), in particular the spin-dependent structure functions,
 not measured so far.

The future of exclusive particle production measurements at charm and 
bottom factories (CLEO-c, BaBar and Belle experiments)
 were also discussed at the 
conference. E.g., at CLEO-c the high statistics and high  precision 
measurements allow progress in measuring branching ratios, form factors, 
decay constants, CKM matrix parameters etc.~\cite{cassel}. 

\section{Conclusion}
Plenty of new high quality data involving photons 
in hadronic  processes have been analysed after PHOTON 2000 at LEP 
and HERA, as well as from  experiments at Tevatron,  RHIC,  VEPP-2M, 
DAPHNE, CESR, KEKB and  SLAC.  
An impressive progress in the measurements is accompanied by
improvements  in the theoretical descriptions of hard, semi-hard 
and soft hadronic processes involving photons. Also pure QED processes
were studied, relevant for present and future colliders.

Processes with photons  provide  simple tests of    new theoretical ideas,
like e.g., non-commutative QED; 
in future conferences of the PHOTON series, we certainly 
will learn more about these new possible interactions of the photon
so that we finally understand  ''what are light quanta''.
\section*{Acknowledgements}
It is pleasure for us to thank the organizers of the conference PHOTON2001, especially Maria Kienzle and Maneesh Wadhwa, for a nice atmosphere and perfect organization in the beautiful site of Ascona. 

Armin B\"ohrer thanks the Bundesministerium f\"ur Bildung und Forschung 
of Germany (grant BMBF 05 HE 1 PS A8) for the financial support. 
Maria Krawczyk is grateful to the organizers of the conference and 
the Polish Committee for Scientific Research (grant 2P03B05119 (2002)) 
for the financial support.

\end{document}